\documentclass[sigconf,10pt]{acmart}

\usepackage[english]{babel}
\usepackage[utf8]{inputenc}
\usepackage{blindtext}
\usepackage{booktabs,caption}
\usepackage{threeparttable}
\usepackage{subfigure}
\usepackage{tumcolor}
\usepackage{balance}
\usepackage{amssymb}
\usepackage{pifont}
\setcopyright{acmcopyright}

\acmDOI{}

\acmISBN{}

\acmConference{ANCS'19}
\acmYear{2019}
\copyrightyear{2019}

\begin{document}

\title{User Space Network Drivers}

\author[Paul Emmerich et al.]{Paul Emmerich}
\affiliation{Technical University of Munich}
\email{emmericp@net.in.tum.de}

\author[ ]{Maximilian Pudelko}
\affiliation{Technical University of Munich}
\affiliation{Open Networking Foundation}
\email{max@opennetworking.org}

\author[ ]{Simon Bauer}
\affiliation{Technical University of Munich}
\email{bauersi@net.in.tum.de}

\author[ ]{Stefan Huber}
\affiliation{Technical University of Munich}
\email{hubestef@net.in.tum.de}

\author[ ]{Thomas Zwickl}
\affiliation{Technical University of Munich}
\email{zwickl@net.in.tum.de}

\author[ ]{Georg Carle}
\affiliation{Technical University of Munich}
\email{carle@net.in.tum.de}

\thispagestyle{empty}

\begin{abstract}
The rise of user space packet processing frameworks like DPDK and netmap makes low-level code more accessible to developers and researchers.
Previously, driver code was hidden in the kernel and rarely modified---or even looked at---by developers working at higher layers.
These barriers are gone nowadays, yet developers still treat user space drivers as black-boxes magically accelerating applications.
We want to change this: every researcher building high-speed network applications should understand the intricacies of the underlying drivers, especially if they impact performance.
We present ixy, a user space network driver designed for simplicity and educational purposes to show that fast packet IO is not black magic but careful engineering.
ixy focuses on the bare essentials of user space packet processing: a packet forwarder including the whole NIC driver uses less than 1,000 lines of C code.

This paper is partially written in tutorial style on the case study of our implementations of drivers for both the Intel 82599 family and for virtual VirtIO NICs.
The former allows us to reason about driver and framework performance on a stripped-down implementation to assess individual optimizations in isolation.
VirtIO support ensures that everyone can run it in a virtual machine.

Our code is available as free and open source under the BSD license at \href{https://github.com/emmericp/ixy}{https://github.com/emmericp/ixy}.
\end{abstract}
\maketitle

\vspace{-7ex}
\section{Introduction}\label{sec:intro}
Low-level packet processing on top of traditional socket APIs is too slow for modern requirements and was therefore often done in the kernel in the past.
Two examples for packet forwarders utilizing kernel components are Open vSwitch~\cite{ovs} and the Click modular router~\cite{click}.
Writing kernel code is not only a relatively cumbersome process with slow turn-around times, it also proved to be too slow for specialized applications. 
Open vSwitch was since extended to include DPDK~\cite{dpdk} as an optional alternative backend to improve performance~\cite{ovs-dpdk}. 
Click was ported to both netmap~\cite{netmap} and DPDK for the same reasons~\cite{fastclick}.
Other projects also moved kernel-based code to specialized user space code~\cite{pfsensedpdk,snortreadme}.

Developers and researchers often treat DPDK as a black-box that magically increases speed.
One reason for this is that DPDK, unlike netmap and others, does not come from an academic background.
It was first developed by Intel and then moved to the Linux Foundation in 2017~\cite{dpdk-linux-foundation}.
This means that there is no academic paper describing its architecture or implementation.
The netmap paper~\cite{netmap} is often used as surrogate to explain how user space packet IO frameworks work in general.
However, DPDK is based on a completely different architecture than seemingly similar frameworks.

Abstractions hiding driver details from developers are an advantage: they remove a burden from the developer.
However, all abstractions are leaky, especially when performance-critical code such as high-speed networking applications are involved.
We therefore believe it is crucial to have at least some insights into the inner workings of drivers when developing high-speed networking applications.

We present ixy, a user space packet framework that is architecturally similar to DPDK~\cite{dpdk} and Snabb~\cite{snabb}.
Both use full user space drivers, unlike netmap~\cite{netmap}, PF\_RING~\cite{pfring}, PFQ~\cite{pfq}, or similar frameworks that rely on a kernel driver.
ixy is designed for educational use only, i.e., you are meant to use it to understand how user space packet frameworks and drivers work, not to use it in a production environment.
Our whole architecture, described in Section~\ref{sec:design}, aims at simplicity and is trimmed down to the bare minimum.
We currently support the Intel ixgbe family of NICs (cf. Section~\ref{sec:impl}) and virtual VirtIO NICs (cf. Section~\ref{sec:virtio}).
A packet forwarding application is less than 1,000 lines of C code including the whole poll-mode driver, the implementation is discussed in Section~\ref{sec:impl}.
It is possible to read and understand drivers found in other frameworks, but ixy's driver is at least an order of magnitude simpler than other implementations.
For example, DPDK's implementation of the ixgbe driver needs 5,400 lines of code just to handle receiving and sending packets in a highly optimized way, offering multiple paths using different vector SIMD instruction sets.
ixy's receive and transmit path for the same driver is only 127 lines of code. 

It is not our goal to support every conceivable scenario, hardware feature, or optimization.
We aim to provide an educational platform for experimentation with driver-level features or optimizations.
ixy is available under the BSD license for this purpose~\cite{ixy}.

\begin{figure}[t]
\vspace{.2cm}
\centering\includegraphics[width=0.40\textwidth]{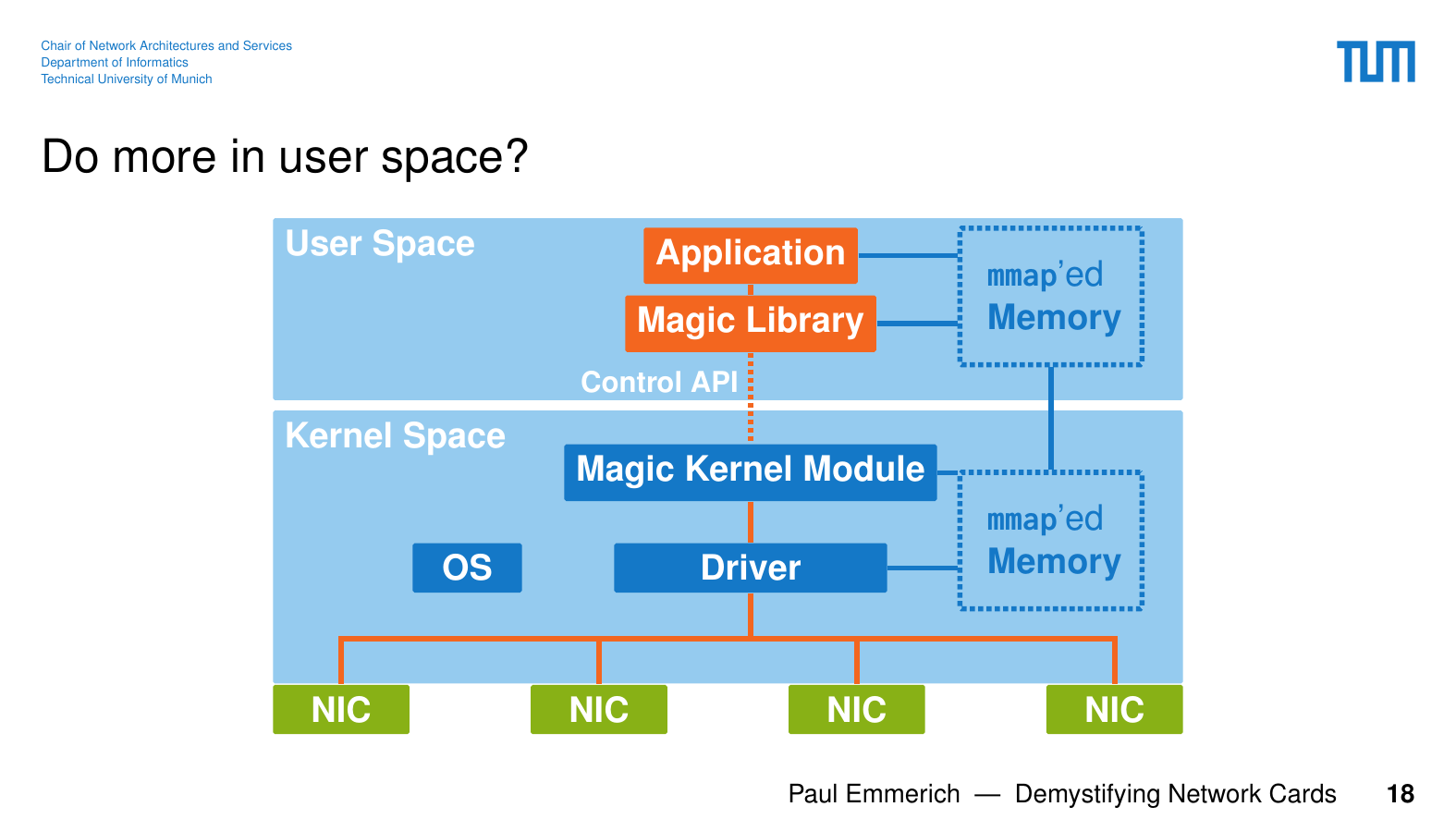}
\caption{Architecture of user space packet processing frameworks using an in-kernel driver, e.g., netmap, PF\_RING ZC, or PFQ.}
\label{fig:driver2}
\vspace{-0.4cm}
\end{figure}

\section{Background and Related Work}
\begin{figure}[t]
\vspace{.2cm}
\centering\includegraphics[width=0.40\textwidth]{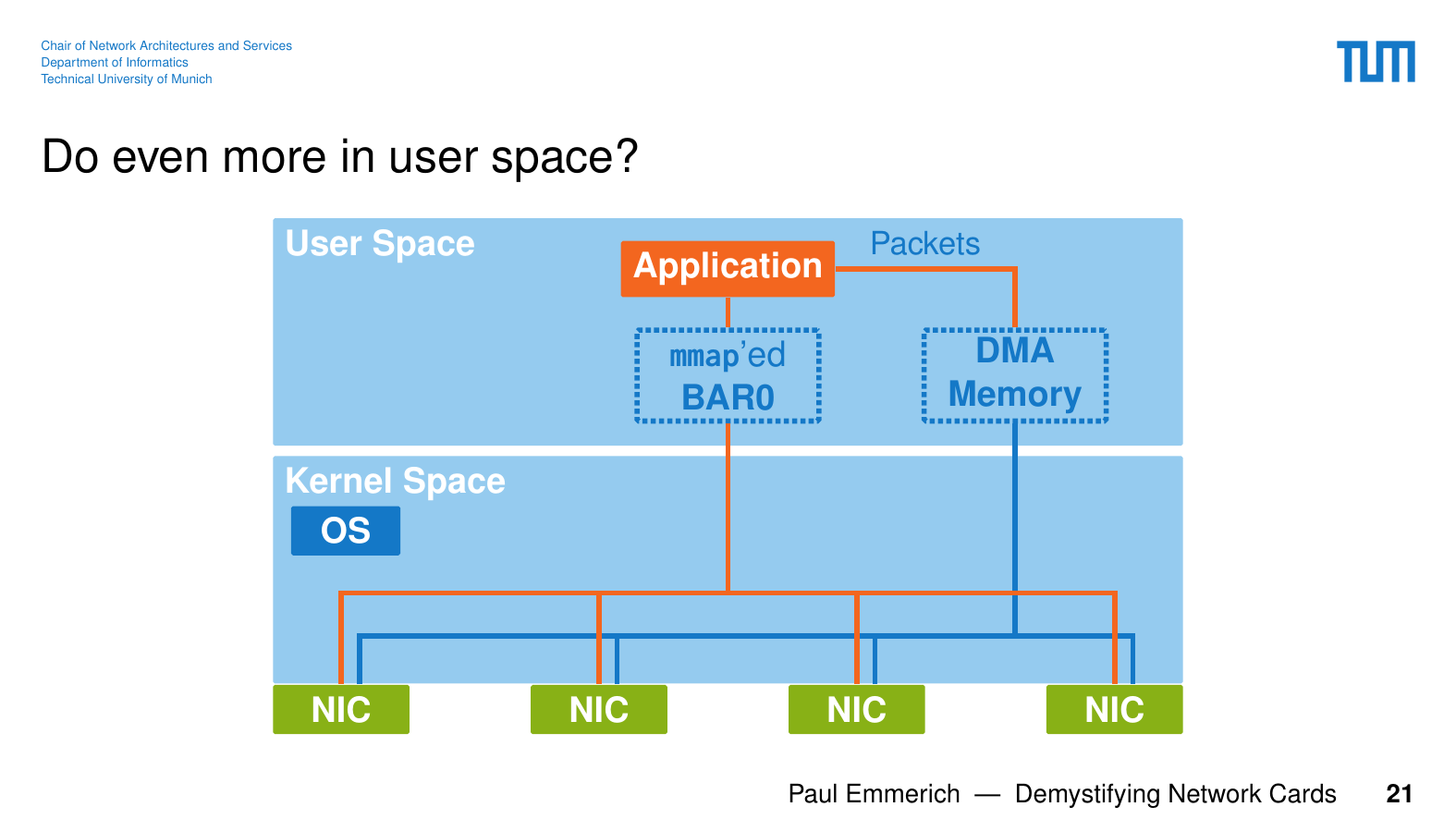}
\caption{Architecture of full user space network drivers, e.g., DPDK, Snabb, or ixy.}
\label{fig:driver3}
\vspace{-0.4cm}
\end{figure}

A multitude of packet IO frameworks have been built over the past years, each focusing on different aspects. 
They can be broadly categorized into two categories: those relying on a driver running in the kernel (Figure~\ref{fig:driver2}) and those that re-implement the whole driver in user space (Figure~\ref{fig:driver3}).

Examples for the former category are netmap~\cite{netmap}, PF\_RING ZC~\cite{pfring}, PFQ~\cite{pfq}, and OpenOnload~\cite{openonload}.
They all use the default driver (sometimes with small custom patches) and an additional kernel component that provides a fast interface based on memory mapping for the user space application.
Packet IO is still handled by the kernel driver here, but the driver is attached to the application directly instead of the kernel datapath, see Figure~\ref{fig:driver2}.
This has the advantage that integrating existing kernel components or forwarding packets to the default network stack is feasible with these frameworks.
By default, these applications still provide an application with exclusive access to the NIC.
Parts of the NIC can often still be controlled with standard tools like \texttt{ethtool} to configure packet filtering or queue sizes.
However, hardware features are often poorly supported, e.g., netmap lacks support for most offloading features~\cite{netmap-offload}.

Note that none of these two advantages is superior to the other, they are simply different approaches for a similar problem. Each solution comes with unique advantages and disadvantages depending on the exact use case.

netmap~\cite{netmap} and XDP~\cite{xdp} are good examples of integrating kernel components with specialized applications.
netmap (a standard component in FreeBSD and also available on Linux) offers interfaces to pass packets between the kernel network stack and a user space app, it can even make use of the kernel's TCP/IP stack with StackMap~\cite{stackmap}.
Further, netmap supports using a NIC with both netmap and the kernel simultaneously by using hardware filters to steer packets to receive queues either managed by netmap or the kernel~\cite{cloudflare-netmap}.
XDP is technically not a user space framework: the code is compiled to eBPF which is run by a JIT in the kernel, this restricts the choice of programming language to those that can target eBPF bytecode (typically a restricted subset of C is used).
It is a default part of the Linux kernel nowadays and hence very well integrated.
It is well-suited to implement firewalls that need to pass on traffic to the network stack~\cite{xdp-ddos}.
More complex applications can be built on top of it with AF\_XDP sockets, resulting in an architecture similar to netmap applications.
Despite being part of the kernel, XDP does not yet work with all drivers as it requires a new memory model for all supported drivers.
At the time of writing, XDP in kernel 4.19 (current LTS) supports fewer drivers than DPDK~\cite{xdp-drivers,dpdk-nics} and does not support forwarding to different NICs.

DPDK~\cite{dpdk}, Snabb~\cite{snabb}, and ixy implement the driver completely in user space.
DPDK still uses a small kernel module with some drivers, but it does not contain driver logic and is only used during initialization. Snabb and ixy require no kernel code at all, see Figure~\ref{fig:driver3}.
A main advantage of the full user space approach is that the application has full control over the driver leading to a far better integration of the application with the driver and the hardware.
DPDK features the largest selection of offloading and filtering features of all investigated frameworks~\cite{dpdk-support-matrix}.
The downside is the poor integration with the kernel, DPDK's KNI (kernel network interface) needs to copy packets to pass them to the kernel unlike XDP or netmap which can just pass a pointer.
Other advantages of DPDK are its support in the industry, mature code base, and large community.
DPDK supports virtually all NICs commonly found in servers~\cite{dpdk-nics}, far more than any other framework we investigated here.

ixy is a full user space driver as we want to explore writing drivers and not interfacing with existing drivers.
Our architecture is based on ideas from both DPDK and Snabb.
The initialization and operation without loading a driver is inspired by Snabb, the API based on explicit memory management, batching, and driver abstraction is similar to DPDK.

\section{Design}
\label{sec:design}
Function names and line numbers referring to our implementation are hyperlinked to the source code on GitHub.

The language of choice for the explanation here and initial implementation is C as the lowest common denominator of systems programming languages.
Implementations in other languages are also available~\cite{ixy-languages-paper}.
Our design goals are:

\begin{itemize}
\item Simplicity. A forwarding application including a driver should be less than 1,000 lines of C code.
\item No dependencies. One self-contained project including the application and driver.
\item Usability. Provide a simple-to-use interface for applications built on it.
\item Speed. It should be reasonable fast without compromising simplicity, find the right trade-off.
\end{itemize}

It should be noted that the Snabb project~\cite{snabb} has similar design goals; ixy tries to be one order of magnitude simpler.
For example, Snabb targets 10,000 lines of code~\cite{snabb-design-goals}, we target 1,000 lines of code and Snabb builds on Lua with LuaJIT instead of C limiting accessibility.

\subsection{Architecture} 
ixy only features one abstraction level: it decouples the used driver from the user's application.
Applications call into ixy to initialize a network device by its PCI address, ixy choses the appropriate driver automatically and returns a struct containing function pointers for driver-specific implementations.
We currently expose packet reception, transmission, and device statistics to the application.
Packet APIs are based on explicit allocation of buffers from specialized \emph{memory pool} data structures.

Applications include the driver directly, ensuring a quick turn-around time when modifying the driver.
This means that the driver logic is only a single function call away from the application logic, allowing the user to read the code from a top-down level without jumping between complex abstraction interfaces or even system calls.

\subsection{NIC Selection}
ixy is based on custom user space re-implementation of the Intel ixgbe driver and the VirtIO virtio-net driver cut down to their bare essentials.
We tested our ixgbe driver on Intel X550, X540, and 82599ES NICs, virtio-net on qemu with and without vhost, and on VirtualBox.
All other frameworks except DPDK are also restricted to very few NIC models (typically 3 or fewer families) and ixgbe is (except for OpenOnload only supporting their own NICs) always supported.

We chose ixgbe for ixy because Intel releases extensive datasheets and the ixgbe NICs are commonly found in commodity servers.
These NICs are also interesting because they expose a relatively low-level interface to the drivers.
Other NICs like the newer Intel XL710 series or Mellanox ConnectX-4/5 follow a more firmware-driven design: a lot of functionality is hidden behind a black-box firmware running on the NIC and the driver merely communicates via a message interface with the firmware which does the hard work.
This approach has obvious advantages such as abstracting hardware details of different NICs allowing for a simpler more generic driver.
However, our goal with ixy is understanding the full stack---a black-box firmware is counterproductive here and we have no plans to add support for such NICs.

VirtIO was selected as second driver to ensure that everyone can run the code without hardware dependencies.
A second interesting characteristic of VirtIO is that it is based on PCI instead of PCIe, requiring a different approach to implement the driver in user space.

\subsection{User Space Drivers in Linux}
There are two subsystems in Linux that enable user space drivers: \texttt{uio} and \texttt{vfio}, we support both.

\texttt{uio} exposes all necessary interfaces to write full user space drivers via memory mapping files in the \texttt{sysfs} pseudo filesystem.
These file-based APIs give us full access to the device without needing to write any kernel code.
ixy unloads any kernel driver for the given PCI device to prevent conflicts, i.e., there is no driver loaded for the NIC while ixy is running.

\texttt{vfio} offers more features: IOMMU and interrupts are only supported with \texttt{vfio}.
However, these features come at the cost of additional complexity:
It requires binding the PCIe device to the generic \texttt{vfio-pci} driver and it then exposes an API via \texttt{ioctl} syscalls on special files.

One needs to understand how a driver communicates with a device to understand how a driver can be written in user space.
A driver can communicate via two ways with a PCIe device:
The driver can initiate an access to the device's Base Address Registers (BARs) or the device can initiate a direct memory access (DMA) to access arbitrary main memory locations.
BARs are used by the device to expose configuration and control registers to the drivers.
These registers are available either via memory mapped IO (MMIO) or via x86 IO ports depending on the device, the latter way of exposing them is deprecated in PCIe.

\subsubsection{Accessing Device Registers}
MMIO maps a memory area to device IO, i.e., reading from or writing to this memory area receives/sends data from/to the device.
\texttt{uio} exposes all BARs in the \texttt{sysfs} pseudo filesystem, a privileged process can simply \texttt{mmap} them into its address space.
\texttt{vfio} provides an \texttt{ioctl} that returns memory mapped to this area.
Devices expose their configuration registers via this interface where normal reads and writes can be used to access registers.
For example, ixgbe NICs expose all configuration, statistics, and debugging registers via the BAR0 address space.
Our implementations of these mappings are in \href{https://github.com/emmericp/ixy/blob/b1cfa2240655f2644f7218abad3141236168f005/src/pci.c#L42}{\texttt{pci\_map\_resource()}} in  \href{https://github.com/emmericp/ixy/blob/b1cfa2240655f2644f7218abad3141236168f005/src/pci.c}{\texttt{pci.c}} and in \href{https://github.com/emmericp/ixy/blob/b1cfa2240655f2644f7218abad3141236168f005/src/libixy-vfio.c#L101}{\texttt{vfio\_map\_region()}} in  \href{https://github.com/emmericp/ixy/blob/b1cfa2240655f2644f7218abad3141236168f005/src/libixy-vfio.c}{\texttt{libixy-vfio.c}}.

VirtIO (in the version we are implementing for compatibility with VirtualBox) is unfortunately based on PCI and not on PCIe and its BAR is an IO port resource that must be accessed with the archaic \texttt{IN} and \texttt{OUT} x86 instructions requiring IO privileges.
Linux can grant processes the necessary privileges via \texttt{ioperm(2)}~\cite{ioperm}, DPDK uses this approach for their VirtIO driver.
We found it too cumbersome to initialize and use as it requires either parsing the PCIe configuration space or text files in \texttt{procfs} and \texttt{sysfs}.
Linux \texttt{uio} also exposes IO port BARs via \texttt{sysfs} as files that, unlike their MMIO counterparts, cannot be \texttt{mmap}ed.
These files can be opened and accessed via normal read and write calls that are then translated to the appropriate IO port commands by the kernel.
We found this easier to use and understand but slower due to the required syscall.
See \href{https://github.com/emmericp/ixy/blob/b1cfa2240655f2644f7218abad3141236168f005/src/pci.c#L54}{\texttt{pci\_open\_resource()}} in \href{https://github.com/emmericp/ixy/blob/b1cfa2240655f2644f7218abad3141236168f005/src/pci.c}{\texttt{pci.c}} and \href{https://github.com/emmericp/ixy/blob/b1cfa2240655f2644f7218abad3141236168f005/src/driver/device.h#L122}{\texttt{read/write\_ioX()}} in \href{https://github.com/emmericp/ixy/blob/b1cfa2240655f2644f7218abad3141236168f005/src/driver/device.h}{\texttt{device.h}} for the implementation.

A potential pitfall is that the exact size of the read and writes are important, e.g., accessing a single 32\,bit register with 2 16\,bit reads will typically fail and trying to read multiple small registers with one read might not be supported.
The exact semantics are up to the device, Intel's ixgbe NICs only expose 32\,bit registers that support partial reads  (except clear-on-read registers) but not partial writes.
VirtIO uses different register sizes and specifies that any access width should work in the mode we are using~\cite{virtiospec}, in practice only aligned and correctly sized accesses work reliably.

\subsubsection{DMA in User Space}\label{sec:dmamemory}
DMA is initiated by the PCIe device and allows it to read/write arbitrary physical addresses.
This is used to access packet data and to transfer the DMA descriptors (pointers to packet data) between driver and NIC.
DMA needs to be explicitly enabled for a device via the PCI configuration space, our implementation is in \href{https://github.com/emmericp/ixy/blob/b1cfa2240655f2644f7218abad3141236168f005/src/pci.c#L27}{\texttt{enable\_dma()}} in \href{https://github.com/emmericp/ixy/blob/b1cfa2240655f2644f7218abad3141236168f005/src/pci.c}{\texttt{pci.c}} for \texttt{uio} and in
\href{https://github.com/emmericp/ixy/blob/b1cfa2240655f2644f7218abad3141236168f005/src/libixy-vfio.c#L21}{\texttt{vfio\_enable\_dma()}} in \href{https://github.com/emmericp/ixy/blob/b1cfa2240655f2644f7218abad3141236168f005/src/libixy-vfio.c}{\texttt{libixy-vfio.c}} for \texttt{vfio}.
DMA memory allocation differs significantly between \texttt{uio} and \texttt{vfio}.

\paragraph{uio DMA memory allocation}

Memory used for DMA transfer must stay resident in physical memory.
\texttt{mlock(2)}~\cite{mlock} can be used to disable swapping.
However, this only guarantees that the page stays backed by memory, it does not guarantee that the physical address of the allocated memory stays the same.
The linux page migration mechanism can change the physical address of any page allocated by the user space at any time, e.g., to implement transparent huge pages and NUMA optimizations~\cite{pagemigration}.
Linux does not implement page migration of explicitly allocated huge pages (2\,MiB or 1\,GiB pages on x86).
ixy therefore uses huge pages which also simplify allocating physically contiguous chunks of memory.
Huge pages allocated in user space are used by all investigated full user space drivers, but they are often passed off as a mere performance improvement~\cite{hugepageperformance1,hugepageperformance2} despite being crucial for reliable allocation of DMA memory.

The user space driver hence also needs to be able to translate its virtual addresses to physical addresses, this is possible via the \texttt{procfs} file \texttt{/proc/self/pagemap}, the translation logic is implemented in \href{https://github.com/emmericp/ixy/blob/b1cfa2240655f2644f7218abad3141236168f005/src/memory.c#L23}{\texttt{virt\_to\_phys()}} in \href{https://github.com/emmericp/ixy/blob/b1cfa2240655f2644f7218abad3141236168f005/src/memory.c}{\texttt{memory.c}}.

\paragraph{vfio DMA memory allocation}
The previous DMA memory allocation scheme is specific to a quirk in Linux on x86 and not portable.
\texttt{vfio} features a portable way to allocate memory that internally calls \texttt{dma\_alloc\_coherent()} in the kernel like an in-kernel driver would.
This syscall abstracts all the messy details and is implemented in our driver in \href{https://github.com/emmericp/ixy/blob/b1cfa2240655f2644f7218abad3141236168f005/src/libixy-vfio.c#L113}{\texttt{vfio\_map\_dma()}} in \href{https://github.com/emmericp/ixy/blob/b1cfa2240655f2644f7218abad3141236168f005/src/libixy-vfio.c}{\texttt{libixy-vfio.c}}.
It requires an IOMMU and configures the necessary mapping to use virtual addresses for the device.

\paragraph{DMA and cache coherency}
Both of our implementations require a CPU architecture with cache-coherent DMA access.
Older CPUs might not support this and require explicit cache flushes to memory before DMA data can be read by the device.
Modern CPUs do not have that problem. In fact, one of the main enabling technologies for high speed packet IO is that DMA accesses do not actually go to memory but to the CPU's cache on any recent CPU architecture. 

\subsubsection{Interrupts in User Space}
\texttt{vfio} features full support for interrupts, \href{https://github.com/emmericp/ixy/blob/f0f2ce884ff6a14cd49d9b9aa1d3518c5a65c180/src/libixy-vfio.c#L150}{\texttt{vfio\_setup\_interrupt()}} in \href{https://github.com/emmericp/ixy/blob/f0f2ce884ff6a14cd49d9b9aa1d3518c5a65c180/src/libixy-vfio.c}{\texttt{libixy-vfio.c}} enables a specific interrupt for vfio and associates it with an eventfd file descriptor.
\href{https://github.com/emmericp/ixy/blob/f0f2ce884ff6a14cd49d9b9aa1d3518c5a65c180/src/driver/ixgbe.c#L160}{\texttt{enable\_msix\_interrupt()}} in \href{https://github.com/emmericp/ixy/blob/f0f2ce884ff6a14cd49d9b9aa1d3518c5a65c180/src/driver/ixgbe.c}{\texttt{ixgbe.c}} configures interrupts for a queue on the device.

Interrupts are mapped to a file descriptor on which the usual syscalls like \texttt{epoll} are available to sleep until an interrupt occurs, see  \href{https://github.com/emmericp/ixy/blob/f0f2ce884ff6a14cd49d9b9aa1d3518c5a65c180/src/libixy-vfio.c#L181}{\texttt{vfio\_epoll\_wait()}} in \href{https://github.com/emmericp/ixy/blob/f0f2ce884ff6a14cd49d9b9aa1d3518c5a65c180/src/libixy-vfio.c}{\texttt{libixy-vfio.c}}.

\subsection{Memory Management}
ixy builds on an API with explicit memory allocation similar to DPDK which is a very different approach from netmap~\cite{netmap} that exposes a replica\footnote{Not the actual ring buffers to prevent user-space applications from crashing the kernel with invalid pointers.}  of the NIC's ring buffer to the application.
Memory allocation for packets was cited as one of the main reasons why netmap is faster than traditional in-kernel processing~\cite{netmap}.
Hence, netmap lets the application handle memory allocation details.
Many forwarding cases can then be implemented by simply swapping pointers in the rings.
However, more complex scenarios where packets are not forwarded immediately to a NIC (e.g., because they are passed to a different core in a pipeline setting) do not map well to this API and require adding manual buffer management on top of this API.
Further, a ring-based API is very cumbersome to use compared to one with memory allocation.

It is true that memory allocation for packets is a significant overhead in the Linux kernel, we have measured a per-packet overhead of 100 cycles\footnote{Forwarding 10 Gbit/s with minimum-sized packets on a single 3.0\,GHz CPU core leaves a budget of 200 cycles/packet.} when forwarding packets with Open vSwitch on Linux for allocating and freeing packet memory (measured with \texttt{perf}).
This overhead is almost completely due to (re-)initialization of the kernel \texttt{sk\_buff} struct: a large data structure with a lot of metadata fields targeted at a general-purpose network stack.
Memory allocation in ixy (with minimum metadata required) only adds an overhead of 30 cycles/packet, a price that we are willing to pay for the gained simplicity in the user-facing API.

ixy's API is the same as DPDK's API when it comes to sending and receiving packets and managing memory.
It can best be explained by reading the example applications \href{https://github.com/emmericp/ixy/blob/b1cfa2240655f2644f7218abad3141236168f005/src/app/ixy-fwd.c}{\texttt{ixy-fwd.c}} and \href{https://github.com/emmericp/ixy/blob/b1cfa2240655f2644f7218abad3141236168f005/src/app/ixy-pktgen.c}{\texttt{ixy-pktgen.c}}.
The transmit-only example \href{https://github.com/emmericp/ixy/blob/b1cfa2240655f2644f7218abad3141236168f005/src/app/ixy-pktgen.c}{\texttt{ixy-pktgen.c}} creates a \emph{memory pool}, a fixed-size collection of fixed-size packet buffers and pre-fills them with packet data.
It then allocates a batch of packets from this pool, adds a sequence number to the packet, and passes them to the transmit function.
The transmit function is asynchronous: it enqueues pointers to these packets, the NIC fetches and sends them later.
Previously sent packets are freed asynchronously in the transmit function by checking the queue for sent packets and returning them to the pool.
This means that a packet buffer cannot be re-used immediately, the \href{https://github.com/emmericp/ixy/blob/b1cfa2240655f2644f7218abad3141236168f005/src/app/ixy-pktgen.c}{\texttt{ixy-pktgen}} example looks therefore quite different from a packet generator built on a classic socket API.

The forward example \href{https://github.com/emmericp/ixy/blob/b1cfa2240655f2644f7218abad3141236168f005/src/app/ixy-fwd.c}{\texttt{ixy-fwd.c}} can avoid explicit handling of memory pools in the application: the driver allocates a memory pool for each receive ring and automatically allocates packets.
Allocation is done by the packet reception function, freeing is either handled in the transmit function as before or by dropping the packet explicitly if the output link is full.
Exposing the rings directly similar to netmap could significantly speed up this simple example application at the cost of usability.

\subsection{Security Considerations}
User space drivers effectively run with root privileges even if they drop privileges after initializing devices: they can use the device's DMA capabilities to access arbitrary memory locations, negating some of the security advantages of running in user space.
This can be mitigated by using the IO memory management unit (IOMMU) to isolate the address space accessible to a device at the cost of an additional (hardware-accelerated) lookup in a page table for each memory access by the device.

IOMMUs are available on CPUs offering hardware virtualization features as they were designed to pass PCIe devices (or parts of them via SR-IOV) directly into VMs in a secure manner.
Linux abstracts different IOMMU implementations via the \texttt{vfio} framework which is specifically designed for ``safe non-privileged userspace drivers''~\cite{vfio} beside virtual machines.
Our \texttt{vfio} backend allows running the driver and application as an unprivileged user.
Of the investigated other frameworks only netmap supports this.
DPDK also offers a \texttt{vfio} backend and has historically supported running with unprivileged users, but recent versions no longer support this with most drivers.
Snabb's \texttt{vfio} backend was removed because of the high maintenance burden and low usage.



\section{ixgbe Implementation}
\label{sec:impl}

All page numbers and section numbers for the Intel datasheet refer to revision 3.3 (March 2016) of the 82599ES datasheet~\cite{82599}.
Function names and line numbers referring to our implementation are hyperlinked to the source code on GitHub.

ixgbe devices expose all configuration, statistics, and debugging registers via the BAR0 MMIO region.
The datasheet lists all registers as offsets in this configuration space in Section 9~\cite{82599}.
We use \texttt{ixgbe\_type.h} from Intel's driver as machine-readable version of the datasheet\footnote{This is technically a violation of both our goals about dependencies and lines of code, but we only effectively use less than 100 lines that are just defines and simple structs. There is nothing to be gained from manually copying offsets and names from the datasheet or this file.}, it contains defines for all register names and offsets for bit fields.

\subsection{NIC Ring API}
NICs expose multiple circular buffers called queues or rings to transfer packets.
The simplest setup uses only one receive and one transmit queue.
Multiple transmit queues are merged on the NIC, incoming traffic is split according to filters or a hashing algorithm if multiple receive queues are configured.
Both receive and transmit rings work in a similar way: the driver programs a physical base address and the size of the ring.
It then fills the memory area with \emph{DMA descriptors}, i.e., pointers to physical addresses where the packet data is stored with some metadata.
Sending and receiving packets is done by passing ownership of the DMA descriptors between driver and hardware via a head and a tail pointer.
The driver controls the tail, the hardware the head. Both pointers are stored in device registers accessible via MMIO.

The initialization code is in \texttt{ixgbe.c} starting from \href{https://github.com/emmericp/ixy/blob/b1cfa2240655f2644f7218abad3141236168f005/src/driver/ixgbe.c#L114}{line 114} for receive queues and from \href{https://github.com/emmericp/ixy/blob/b1cfa2240655f2644f7218abad3141236168f005/src/driver/ixgbe.c#L173}{line 173} for transmit queues.
Further details are in the datasheet in Section 7.1.9 and in the datasheet sections mentioned in the code.

\subsubsection{Receiving Packets}

\begin{figure}[t]
\centering\includegraphics[width=0.44\textwidth]{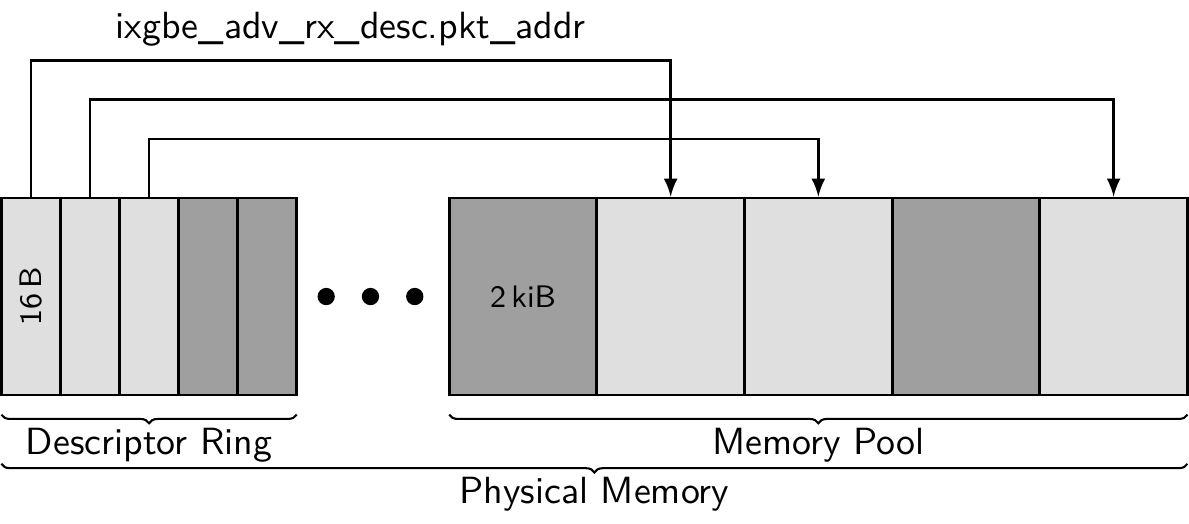}
\vspace{-0.2cm}
\caption{DMA descriptors pointing into a memory pool, note that the packets in the memory are unordered as they can be free'd at different times.}
\label{fig:memory}
\vspace{0.2cm}
\end{figure}

The driver fills up the ring buffer with physical pointers to packet buffers in \href{https://github.com/emmericp/ixy/blob/b1cfa2240655f2644f7218abad3141236168f005/src/driver/ixgbe.c#L64}{\texttt{start\_rx\_queue()}} on startup.
Each time a packet is received, the corresponding buffer is returned to the application and we allocate a new packet buffer and store its physical address in the DMA descriptor and reset the ready flag.
We also need a way to translate the physical addresses in the DMA descriptor found in the ring back to its virtual counterpart on packet reception.
This is done by keeping a second copy of the ring populated with virtual instead of physical addresses, this is then used as a lookup table for the translation.

\begin{figure}[t]
\hspace{-0.1cm}\includegraphics[width=0.45\textwidth]{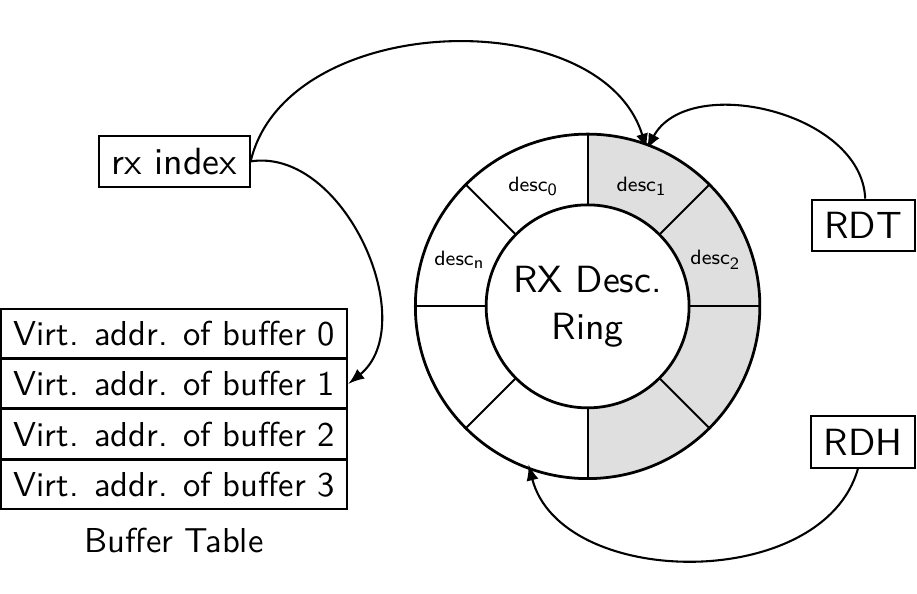}
\caption{Overview of a receive queue. The ring uses physical addresses and is shared with the NIC.}
\label{fig:rx-ring}
\vspace{-.4cm}
\end{figure}

Figure~\ref{fig:memory} illustrates the memory layout: the DMA descriptors in the ring to the left contain physical pointers to packet buffers stored in a separate location in a memory pool.
The packet buffers in the memory pool contain their physical address in a metadata field.
Figure~\ref{fig:rx-ring} shows the RDH (head) and RDT (tail) registers controlling the ring buffer on the right side, and the local copy containing the virtual addresses to translate the physical addresses in the descriptors in the ring back for the application.
\href{https://github.com/emmericp/ixy/blob/b1cfa2240655f2644f7218abad3141236168f005/src/driver/ixgbe.c#L389}{\texttt{ixgbe\_rx\_batch()}} in \texttt{ixgbe.c} implements the receive logic as described by Sections 1.8.2 and 7.1 of the datasheet.
It operates on batches of packets to increase performance.
A na\"ive way to check if packets have been received is reading the head register from the NIC incurring a PCIe round trip.
The hardware also sets a flag in the descriptor via DMA which is far cheaper to read as the DMA write is handled by the last-level cache on modern CPUs.
This is effectively the difference between an LLC cache miss and hit for every received packet.

\subsubsection{Transmitting Packets}
Transmitting packets follows the same concept and API as receiving them, but the function is more complicated because the interface between NIC and driver is asynchronous.
Placing a packet into the ring does not immediately transfer it and blocking to wait for the transfer is infeasible.
Hence, the \href{https://github.com/emmericp/ixy/blob/b1cfa2240655f2644f7218abad3141236168f005/src/driver/ixgbe.c#L442}{\texttt{ixgbe\_tx\_batch()}} function in \texttt{ixgbe.c} consists of two parts: freeing packets from previous calls that were sent out by the NIC followed by placing the current packets into the ring.
The first part is often called cleaning and works similar to receiving packets: the driver checks a flag that is set by the hardware after the packet associated with the descriptor is sent out.
Sent packet buffers can then be free'd, making space in the ring.
Afterwards, the pointers of the packets to be sent are stored in the DMA descriptors and the tail pointer is updated accordingly.

Checking for transmitted packets can be a bottleneck due to cache thrashing as both the device and driver access the same memory locations~\cite{82599}.
The 82599 hardware implements two methods to combat this: marking transmitted packets in memory occurs either automatically in configurable batches on device side, this can also avoid unnecessary PCIe transfers.
We tried different configurations (code in \href{https://github.com/emmericp/ixy/blob/b1cfa2240655f2644f7218abad3141236168f005/src/driver/ixgbe.c#L173}{\texttt{init\_tx()}}) and found that the defaults from Intel's driver work best.
The NIC can also write its current position in the transmit ring back to memory periodically (called head pointer write back) as explained in Section~7.2.3.5.2 of the datasheet.
However, no other driver implements this feature despite the datasheet referring to the normal marking mechanism as ``legacy''.
We implemented support for head pointer write back on a branch~\cite{ixy-headptr-writeback} but found no measurable performance improvements or effects on cache contention.

\subsubsection{Batching}
Each successful transmit or receive operation involves an update to the NIC's tail pointer register (RDT or TDT for receive/transmit), a slow operation.
This is one of the reasons why batching is so important for performance.
Both the receive and transmit function are batched in ixy, updating the register only once per batch.

\subsubsection{Offloading Features}
ixy currently only enables CRC checksum offloading.
Unfortunately, packet IO frameworks (e.g., netmap) are often restricted to this bare minimum of offloading features.
DPDK is the exception here as it supports almost all offloading features offered by the hardware.
However, its receive and transmit functions pay the price for these features in the form of complexity.

We will try to find a balance and showcase selected simple offloading features in ixy in the future.
These offloading features can be implemented in the receive and transmit functions, see comments in the code.
This is simple for some features like VLAN tag offloading and more involved for more complex features requiring an additional descriptor containing metadata information.

\begin{figure}[t]
\includegraphics[width=0.47\textwidth]{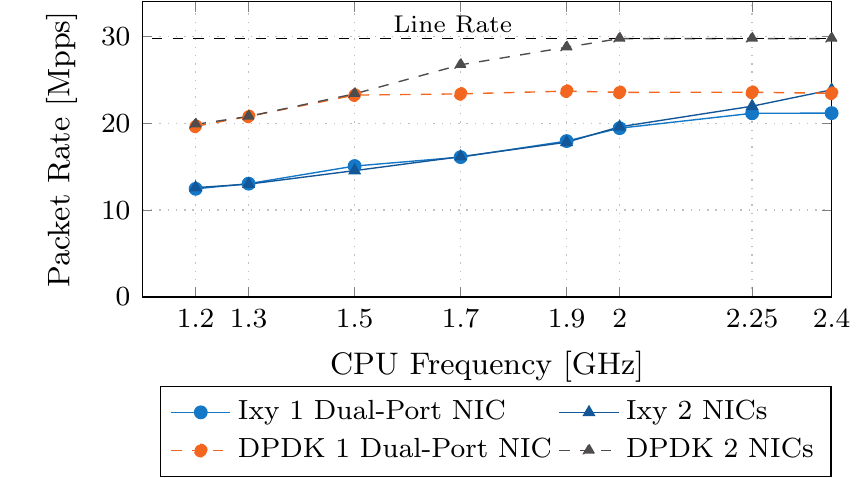}
\vspace{-0.3em}
\caption{Bidirectional single-core forwarding performance with varying CPU speed, batch size 32.}
\label{fig:freq}
\end{figure}

\section{Performance Evaluation}
\label{sec:perf}
We run the \texttt{ixy-fwd} example under a full bidirectional load of 29.76 million packets per second (Mpps), line rate with minimum-sized packets at 2x 10\,Gbit/s, and compare it to a custom DPDK forwarder implementing the same features.
Both forwarders modify a byte in the packet to ensure that the packet data is fetched into the L1 cache to simulate a somewhat realistic workload.

\subsection{Throughput}
To quantify the baseline performance and identify bottlenecks, we run the forwarding example while increasing the CPU's clock frequency from 1.2\,GHz to 2.4\,GHz.
Figure~\ref{fig:freq} compares the throughput of our forwarder on ixy and on DPDK when forwarding across the two ports of a dual-port NIC and when using two separate NICs.
The better performance of both ixy and DPDK when using two separate NICs over one dual-port NIC indicates a hardware limit (likely at the PCIe level).
We run this test on Intel X520 (82599-based) and Intel X540 NICs with identical results.
ixy requires 96 CPU cycles to forward a packet, DPDK only 61.
The high performance of DPDK can be attributed to its \emph{vector transmit path} utilizing SIMD instructions to handle batches even better than ixy.
This transmit path of DPDK is only used if no offloading features are enabled at device configuration time, i.e., it offers a similar feature set to ixy.
Disabling the vector TX path in the DPDK configuration, or using an older version of DPDK, increases the CPU cycles per packet to 91 cycles packet, still slightly faster than ixy despite doing more (checking for more offloading flags).
Overall, we consider ixy fast enough for our purposes.
For comparison, performance evaluations of older (2015) versions of DPDK, PF\_RING, and netmap and required $\approx$100 cycles/packet for DPDK and PF\_RING and $\approx$120 cycles/packet for netmap~\cite{gallenmueller2015comparison}.

\begin{figure}[t]
\includegraphics[width=0.47\textwidth]{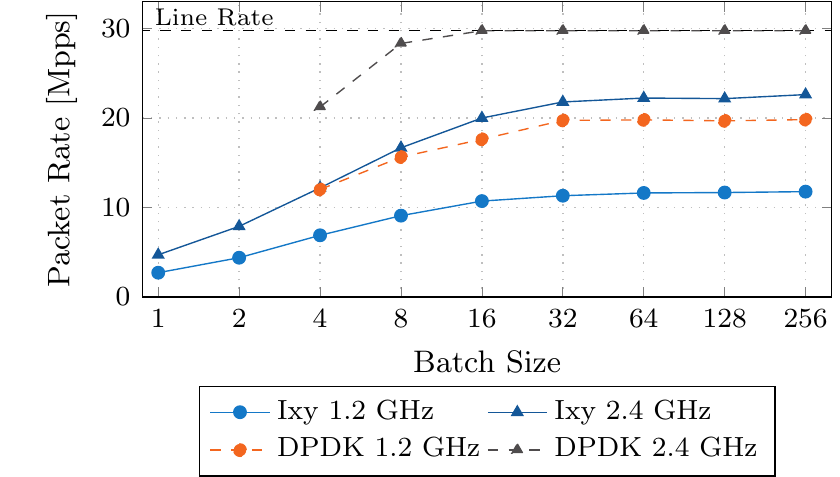}
\vspace{-0.3em}
\caption{Bidirectional single-core forwarding performance with varying batch size.}
\vspace{-0.5cm}
\label{fig:batches}
\end{figure}

\subsection{Batching}
Batching is one of the main drivers for performance. DPDK even requires a minimum batch size of 4 when using the SIMD transmit path.
Receiving or sending a packet involves an access to the queue index registers, invoking a costly PCIe round-trip.
Figure~\ref{fig:batches} shows how the performance increases as the batch size is increased in the bidirectional forwarding scenario with two NICs.
Increasing batch sizes have diminishing returns: this is especially visible when the CPU is only clocked at 1.2\,GHz.
Reading the performance counters for all caches shows that the number of L1 cache misses per packet increases as the performance gains drop off.
Too large batches thrash the L1 cache, possibly evicting lookup data structures in a real application. Therefore, batch sizes should not be chosen too large.
Latency is also impacted by the batch size, but the effect is negligible compared to other buffers (e.g., NIC ring size of 512).

\begin{table}[t]
 \setlength{\tabcolsep}{1mm}
	\centering
	\footnotesize
	\begin{tabular}{lrrrrr}
		Intr./polling & Load & Median & 99th perc. & 99.99th perc. & Max\\
		\toprule
		Polling & 0.1\,Mpps & 3.8\,$\mu$s & 4.7\,$\mu$s & 5.8\,$\mu$s & 15.8\,$\mu$s\\
		Intr., no throttling & 0.1\,Mpps & 7.7\,$\mu$s & 8.4\,$\mu$s & 11.0\,$\mu$s & 76.6\,$\mu$s\\
		Intr., ITR 10\,$\mu$s & 0.1\,Mpps & 11.3\,$\mu$s & 11.9\,$\mu$s & 15.3\,$\mu$s & 78.4\,$\mu$s\\
		Intr., ITR 200\,$\mu$s & 0.1\,Mpps & 107.4\,$\mu$s & 208.1\,$\mu$s & 240.0\,$\mu$s & 360.0\,$\mu$s \vspace{0.5em}\\
		Polling & 0.4\,Mpps & 3.8\,$\mu$s & 4.6\,$\mu$s & 5.8\,$\mu$s & 16.4\,$\mu$s\\
		Intr., no throttling & 0.4\,Mpps & 7.3\,$\mu$s & 8.2\,$\mu$s & 10.9\,$\mu$s & 53.9\,$\mu$s\\
		Intr., ITR 10\,$\mu$s & 0.4\,Mpps & 9.0\,$\mu$s & 14.0\,$\mu$s & 25.8\,$\mu$s & 86.1\,$\mu$s \\
		Intr., ITR 200\,$\mu$s & 0.4\,Mpps & 105.9\,$\mu$s & 204.6\,$\mu$s & 236.7\,$\mu$s & 316.2\,$\mu$s \vspace{0.5em}\\
		Polling & 0.8\,Mpps & 3.8\,$\mu$s & 4.4\,$\mu$s & 5.6\,$\mu$s & 16.8\,$\mu$s\\
		Intr., no throttling & 0.8\,Mpps & 5.6\,$\mu$s & 8.2\,$\mu$s & 10.8\,$\mu$s & 81.1\,$\mu$s\\
		Intr., ITR 10\,$\mu$s & 0.8\,Mpps & 9.2\,$\mu$s & 14.1\,$\mu$s & 31.0\,$\mu$s & 70.2\,$\mu$s  \\
		Intr., ITR 200\,$\mu$s & 0.8\,Mpps & 102.8\,$\mu$s & 198.8\,$\mu$s & 226.1\,$\mu$s & 346.8\,$\mu$s \\
		\bottomrule
	\end{tabular}
		\caption{Forwarding latency by interrupt/poll mode.}
		\vspace{-1.4em}
	\label{tbl:interrupts}
\end{table}

\subsection{Interrupts}
Interrupts are a common mechanism to reduce power consumption at low loads.
However, interrupts are expensive: they require multiple context switches. This makes them unsuitable for high packet rates.
NICs commonly feature interrupt throttling (ITR, configured in $\mu$s between interrupts on the NIC used here) to prevent overloading the system. 
Operating systems often disable interrupts periodically and switch to polling during periods of high loads (e.g., Linux NAPI).
Our forwarder loses packets in interrupt mode at rates of above around 1.5\,Mpps even with aggressive throttling configured on the NIC.
All other tests except this one are therefore conducted in pure polling mode.

A common misconception is that interrupts reduce latency, but they actually increase latency.
The reason is that an interrupt first needs to wake the system from sleep (sleep states down to C6 are enabled on the test system), trigger a context switch into interrupt context, trigger another switch to the driver and then poll the packets from the NIC\footnote{The same is true for Linux kernel drivers, the actual packet reception is not done in the hardware interrupt context but in a software interrupt}.
Permanently polling for new packets in a busy-wait loop avoids this at the cost of power consumption.

Table~\ref{tbl:interrupts} shows latencies at low rates (where interrupts are effective) with and without interrupt throttling (ITR) and polling.
Especially tail latencies are affected by using interrupts instead of polling.
All timestamps were acquired with a fiber-optic splitter and a dedicated timestamping device taking timestamps of every single packet.

These results show that interrupts with a low throttle rate are feasible at low packet rates. 
Interrupts are poorly supported in other user space drivers: Snabb offers no interrupts, DPDK has limited support for interrupts (only some drivers) without built-in automatic switching between different modes.
Frameworks relying on a kernel driver can use the default driver's interrupt features, especially netmap offers good support for power-saving via interrupts.


\begin{table}[t]
 \setlength{\tabcolsep}{1.5mm}
	\centering
	\footnotesize
	\begin{tabular}{lrrrr}
		\textbf{App/Function} & RX & TX & Forwarding & Memory Mgmt.\\
		\toprule
		\textbf{ixy-fwd} & 44.8 & 14.7 & 12.3 & 30.4\\
		\textbf{ixy-fwd-inline} & 57.0 & 28.3 & 12.5 & ?$^*$\\
		\textbf{DPDK l2fwd} &  35.4 & 20.4 & $^\dagger$6.1 & ?$^*$\\
		\textbf{DPDK v1.6 l2fwd$^\ddagger$} & 41.7 & 53.7 & $^\dagger$6.0 & ?$^*$\\
		\bottomrule
	\end{tabular}
        \begin{tablenotes}
            \item \hspace{-1.5em} $^*$\footnotesize Memory operations inlined, separate profiling not possible.
            \item \hspace{-1.5em} $^\dagger$\footnotesize DPDK's driver explicitly prefetches packet data on RX, so this is faster despite performing the same action of changing one byte.
            \item \hspace{-1.5em} $^\ddagger$\footnotesize Old version 1.6 (2014) of DPDK, far fewer SIMD optimizations, measured on a different system/kernel due to compatibility.
        \end{tablenotes}
		\caption{Processing time in cycles per packet.}
		\vspace{-1.4em}
	\label{tbl:perf}
\end{table}

\subsection{Profiling}
We run \texttt{perf} on \texttt{ixy-fwd} running under full bidirectional load at 1.2\,GHz with two different NICs using the default batch size of 32 to ensure that the CPU is the only bottleneck.
\texttt{perf} allows profiling with the minimum possible effect on the performance: throughput drops by only $\approx$5\% while \texttt{perf} is running.
Table~\ref{tbl:perf} shows where CPU time is spent on average per forwarded packet and compares it to DPDK.
Receiving is slower because the receive logic performs the initial fetch, the following functions operate on the L1 cache.
ixy's receive function still leaves room for improvements, it is less optimized than the transmit function.
There are several places in the receive function where DPDK avoids memory accesses by batching compared to ixy.
However, these optimizations were not applied for simplicity in ixy: DPDK's receive function is quite complex and full of SIMD intrinsics leading to poor readability.
We also compare an old version of DPDK in the table that did not yet contain as many optimizations; ixy outperforms the old DPDK version at low CPU speeds, but the old DPDK version is $\approx$10\% faster than ixy at higher CPU speeds indicating better utilization of the CPU pipeline.

Overhead for memory management is significant (but still low compared to the 100 cycles/packet we measured in the Linux kernel).
59\% of the time is spent in non-batched memory operations and none of the calls are inlined.
Inlining these functions increases throughput by 6.5\% but takes away our ability to account time spent in them. 
Overall, the overhead of memory management is larger than we initially expected, but we still think explicit memory management for the sake of a usable API is a good trade-off.
This is especially true for ixy aiming at simplicity, but also for other frameworks targeting complex applications.
Simple forwarding can easily be done on an exposed ring interface, but anything more complex that does not sent out packets immediately (e.g., because they are processed further on a different core) requires memory management in the user's application with a similar performance impact.

\begin{table}[b]
 \setlength{\tabcolsep}{1.4mm}
	\centering
	\footnotesize
	\begin{tabular}{lrrrr}
		Ring Sizes & Load & Median & 99th perc. & 99.9th perc.\\
		\toprule
		64 & 15\,Mpps & 5.2\,$\mu$s & 6.4\,$\mu$s & 7.2\,$\mu$s\\
		512 & 15\,Mpps & 5.2\,$\mu$s & 6.5\,$\mu$s & 7.5\,$\mu$s\\
		4096 & 15\,Mpps & 5.4\,$\mu$s & 6.8\,$\mu$s & 8.7\,$\mu$s\\
		64 & $^*$29\,Mpps & 8.3\,$\mu$s & 9.1\,$\mu$s & 10.6\,$\mu$s\\
		512 & $^*$29\,Mpps & 50.9\,$\mu$s & 52.3\,$\mu$s & 54.3\,$\mu$s\\
		4096 & $^*$29\,Mpps & 424.7\,$\mu$s & 433.0\,$\mu$s & 442.1\,$\mu$s\\
		\bottomrule
	\end{tabular}
        \begin{tablenotes}
            \item \hspace{-1.5em} $^*$\footnotesize Device under test overloaded, packets were lost
        \end{tablenotes}
		\caption{Forwarding latency by ring size and load.}
	\label{tbl:ring-lat}
\end{table}

\subsection{Queue Sizes}
\begin{figure}[t]
\includegraphics[width=0.47\textwidth]{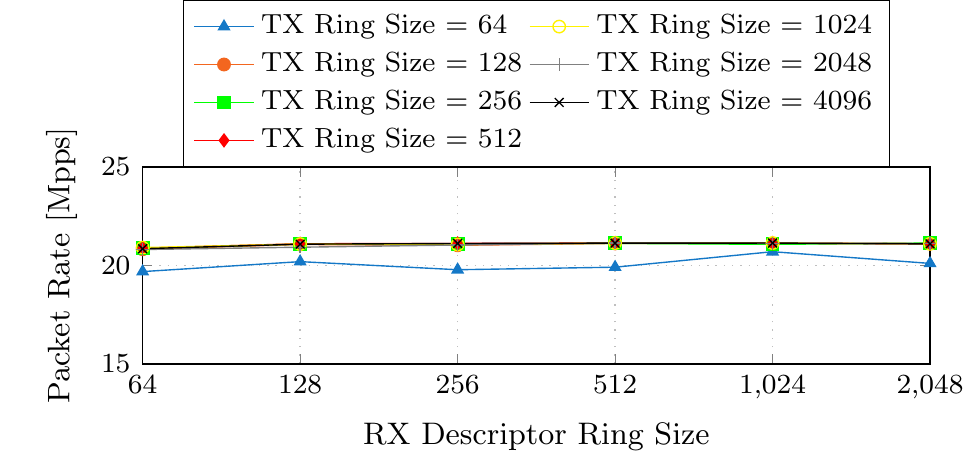}
\vspace{-0.3em}
\caption{Throughput with varying descriptor ring sizes at 2.4\,GHz.}
\vspace{-0.75em}
\label{fig:rings}
\end{figure}

Our driver supports descriptor ring sizes in power-of-two increments between 64 and 4096, the hardware supports more sizes but the restriction to powers of two simplifies wrap-around handling.
Linux defaults to a ring size of 256 for this NIC, DPDK's example applications configure different sizes; the \texttt{l2fwd} forwarder sets 128/512 RX/TX descriptors. 
Larger ring sizes such as 8192 are sometimes recommended to increase performance~\cite{redhattuning} (source refers to the size as kB when it is actually number of packets).
Figure~\ref{fig:rings} shows the throughput of ixy with various ring size combinations.
There is no measurable impact on the maximum throughput for ring sizes larger than 64.
Scenarios where a larger ring size can still be beneficial might exist: for example, an application producing a large burst of packets significantly faster than the NIC can handle for a very short time.

The second performance factor that is impacted by ring sizes is the overall latency caused by unnecessary buffering.
Table~\ref{tbl:ring-lat} shows the latency (measured with MoonGen hardware timestamping~\cite{moongen}) of the ixy forwarder with different ring sizes.
The results show a linear dependency between ring size and latency when the system is overloaded, but the effect under lower loads are negligible.
Full or near full buffers are no exception on systems forwarding Internet traffic due to protocols like TCP that try to fill up buffers completely~\cite{gettys2011bufferbloat}.
We conclude that tuning tips like setting ring sizes to 8192~\cite{redhattuning} are detrimental for latency and likely do not help with throughput.
ixy uses a default ring size of 512 at the moment as a trade-off between providing some buffer and avoiding high worst-case latencies.

\subsection{Page Sizes without IOMMU}
\begin{figure}[t]
\includegraphics[width=0.47\textwidth]{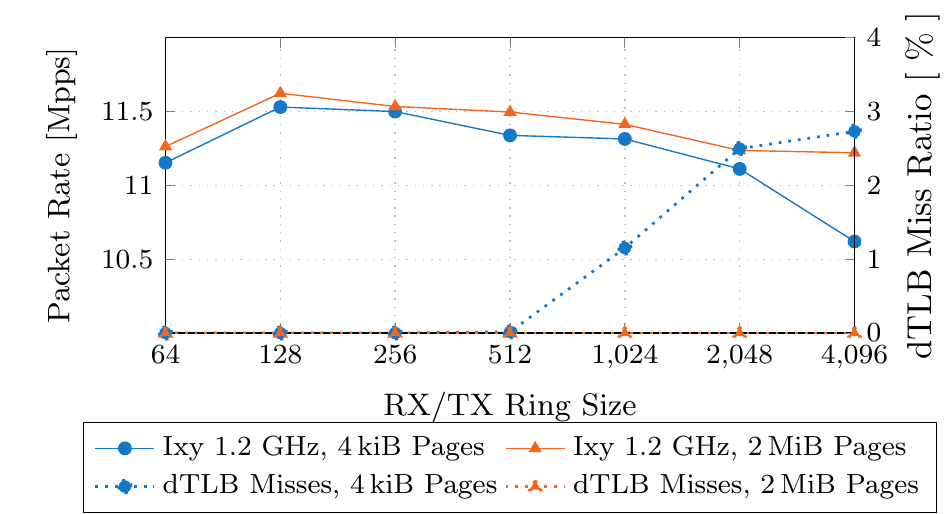}
\vspace{-0.3em}
\caption{Single-core forwarding performance with and without huge pages and their effect on the TLB.}
\label{fig:small-pages}
\end{figure}

It is not possible to allocate DMA memory on small pages from user space in Linux in a reliable manner without using the IOMMU  as described in Section~\ref{sec:dmamemory}.
Despite this, we have implemented an allocator that performs a brute-force search for physically contiguous normal-sized pages from user space.
We run this code on a system without NUMA and with transparent huge pages and page-merging disabled to avoid unexpected page migrations.
The code for these benchmarks is hidden on a branch~\cite{ixy-smallpage} due to its unsafe nature on some systems (we did lose a file system to rogue DMA writes on a misconfigured server).
Benchmarks varying the page size are interesting despite these problems: kernel drivers and user space packet IO frameworks using kernel drivers only support normal-sized pages.
Existing performance claims about huge pages in drivers are vague and unsubstantiated~\cite{hugepageperformance1,hugepageperformance2}.

Figure~\ref{fig:small-pages} shows that the impact on performance of huge pages in the driver is small.
The performance difference is 5.5\% with the maximum ring size, more realistic ring sizes only differ by 1-3\%.
This is not entirely unexpected: the largest queue size of 4096 entries is only 16\,kiB large, storing pointers to up to 16\,MiB packet buffers.
Huge pages are designed for, and usually used with, large data structures, e.g., big lookup tables for forwarding.
The effect measured here is likely larger when a real forwarding application puts additional pressure on the TLB (4096 entries on the CPU used here) due to its other internal data structures.
One should still use huge pages for other data structures in a packet processing application, but a driver not supporting them is not as bad as one might expect when reading claims about their importance from authors of drivers supporting them.
\begin{figure}[t!]
     \includegraphics[width=0.45\textwidth]{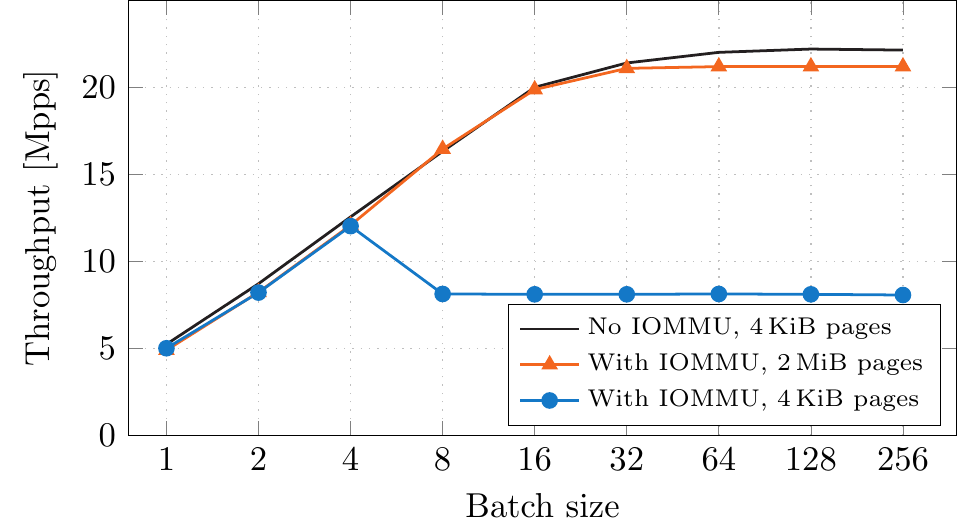}
    \caption{IOMMU impact on single-core forwarding at 2.4\,GHz.}
    \label{fig:iommu}
\end{figure}

\subsection{Page Sizes and IOMMU Overhead}
Memory access overhead changes if the device has to go through the IOMMU for every access.
Documentation for Intel's IOMMU is sparse: The TLB size is not documented and there are no dedicated performance counters.
Neugebauer et al. experimentally determined a TLB size of 64 entries with their pcie-bench framework~\cite{pciebench} (vs. 4096 entries in the normal TLB).
They note a performance impact for small DMA transactions with large window sizes: 64\,byte read throughput drops by up to 70\%, write throughput by up to 55\%.
256\,byte reads are 30\% slower, only 512\,byte and larger transactions are unaffected~\cite{pciebench}.
Their results are consistent across four different Intel microarchitectures including the CPU we are using here.
They explicitly disable huge pages for their IOMMU benchmark.

Our benchmark triggers a similar scenario when not used with huge pages: We ask the NIC to transfer a large number of small packets via DMA.
Note that packets in a batch are not necessarily contiguous in memory: Buffers are not necessarily allocated sequentially and each DMA buffer is 2\,kiB large by default, of which only the first $n$ bytes will be transferred.
This means only two packets share a 4\,kiB page, even if the packets are small.
2\,kiB is a common default in other drivers as it allows handling normal sized frames without chaining several buffers (the NIC only supports DMA buffers that are a multiple of 1\,kiB).
The NIC therefore has to perform several small DMA transactions, i.e., the scenario is equivalent to a large transfer window in pcie-bench.

Figure~\ref{fig:iommu} shows that the IOMMU does not affect the performance if used with 2\,MiB pages.
However, the default 4\,KiB pages (that are safe and easy to use with \texttt{vfio} and the IOMMU) are affected by the small TLB in the IOMMU.
The impact of the IOMMU on our real application is slightly smaller than in the synthetic pcie-bench tests:
The IOMMU costs 62\% performance for the commonly used batch size of 32 with small packets when not using huge pages.
Running the test with 128\,byte packets causes 33\% performance loss, 256\,byte packets yield identical performance.

However, enabling huge pages completely mitigates the impact of the small TLB in the IOMMU.
Note that huge pages for the IOMMU are only supported since the Intel Ivy Bridge CPU generation.


\subsection{NUMA Considerations}
Non-uniform memory access (NUMA) architectures found on multi-CPU servers present additional challenges.
Modern systems integrate cache, memory controller, and PCIe root complex in the CPU itself instead of using a separate IO hub.
This means that a PCIe device is attached to only one CPU in a multi-CPU system, access from or to other CPUs needs to pass over the CPU interconnect (QPI on our system).
At the same time, the tight integration of these components allows the PCIe controller to transparently write DMA data into the cache instead of main memory.
This works even when DCA (direct cache access) is not used (DCA is only supported by the kernel driver, none of the full user space drivers implement it).
Intel DDIO (Data Direct I/O) is another optimization to prevent memory accesses by DMA~\cite{ddio}.
However, we found by reading performance counters that even CPUs not supporting DDIO do not perform memory accesses in a typical packet forwarding scenario.
DDIO is poorly documented and exposes no performance counters, its exact effect on modern systems is unclear.
All recent (since 2012) CPUs supporting multi-CPU systems also support DDIO.
Our NUMA benchmarks where obtained on a different system than the previous results because we want to avoid potential problems with NUMA for the other setups.

\begin{table}[t]
 \setlength{\tabcolsep}{1.15mm}
 	\footnotesize
	\centering
	\begin{tabular}{lrrrr}
		Ingress$^*$ & Egress$^*$ & CPU$^\dagger$ & Memory$^\ddagger$ & Throughput\\
		\toprule
		Node 0 & Node 0 & Node 0 & Node 0 & 10.8\,Mpps\\
		Node 0 & Node 0 & Node 0 & Node 1 & 10.8\,Mpps\\
		Node 0 & Node 0 & Node 1 & Node 0 & 7.6\,Mpps\\
		Node 0 & Node 0 & Node 1 & Node 1 & 6.6\,Mpps\\
		Node 0 & Node 1 & Node 0 & Node 0 & 7.9\,Mpps\\
		Node 0 & Node 1 & Node 0 & Node 1 & 10.0\,Mpps\\
		Node 0 & Node 1 & Node 1 & Node 0 & 8.6\,Mpps\\
		Node 0 & Node 1 & Node 1 & Node 1 & 8.1\,Mpps\\
		\bottomrule
	\end{tabular}
        \begin{tablenotes}
		\item \hspace{-1.5em} $^*$\footnotesize NUMA node connected to the NIC
		\item \hspace{-1.5em} $^\dagger$\footnotesize Thread pinned to this NUMA node
		\item \hspace{-1.5em} $^\ddagger$\footnotesize Memory pinned to this NUMA node
        \end{tablenotes}
		\caption{Unidirectional forwarding on a NUMA system, both CPUs at 1.2 GHz.}
	\label{tbl:numa}
	\vspace{-2em}
\end{table}

Our test system has one dual-port NIC attached to NUMA node 0 and a second to NUMA node 1.
Both the forwarding process and the memory used for the DMA descriptors and packet buffers can be explicitly pinned to a NUMA node.
This gives us 8 possible scenarios for unidirectional packet forwarding by varying the packet path and pinning.
Table~\ref{tbl:numa} shows the throughput at 1.2\,GHz.
Forwarding from and to a NIC at the same node shows one unexpected result: pinning memory, but not the process itself, to the wrong NUMA node does not reduce performance.
The explanation for this is that the DMA transfer is still handled by the correct NUMA node to which the NIC is attached, the CPU then caches this data while informing the other node.
However, the CPU at the other node never accesses this data and there is hence no performance penalty.
Forwarding between two different nodes is fastest when the the memory is pinned to the egress nodes and CPU to the ingress node and slowest when both are pinned to the ingress node.
Real forwarding applications often cannot know the destination of packets at the time they are received, the best guess is therefore to pin the thread to the node local to the ingress NIC and distribute packet buffer across the nodes. 
Latency was also impacted by poor NUMA mapping, we measured an additional 1.7\,$\mu$s when unnecessarily crossing the NUMA boundary.


\section{VirtIO Implementation}\label{sec:virtio}
All section numbers for the specification refer to version 1.0 of the VirtIO specification~\cite{82599}.
Function names and line numbers referring to our implementation are hyperlinked to the source code on GitHub.

VirtIO defines different types of operational modes for emulated network cards: legacy, modern, and transitional devices.
qemu implements all three modes, the default being transitional devices supporting both the legacy and modern interface after feature negotiation.
Supporting devices operating only in modern mode would be the simplest implementation in ixy because they work with MMIO.
Both legacy and transitional devices require support for PCI IO port resources making the device access different from the ixgbe driver.
Modern-only devices are rare because they are relatively new (2016).

We chose to implement the legacy variant as VirtualBox only supports the legacy operation mode.
VirtualBox is an important target as it is the only hypervisor supporting VirtIO that is available on all common operating systems.
Moreover, it is very well integrated with Vagrant~\cite{vagrant} allowing us to offer a self-contained setup to run ixy on any platform~\cite{ixy-vagrant}.

\subsection{Device Initialization and Virtqueues}
\begin{figure}[t]
\hspace{-2mm}\includegraphics[width=0.46\textwidth]{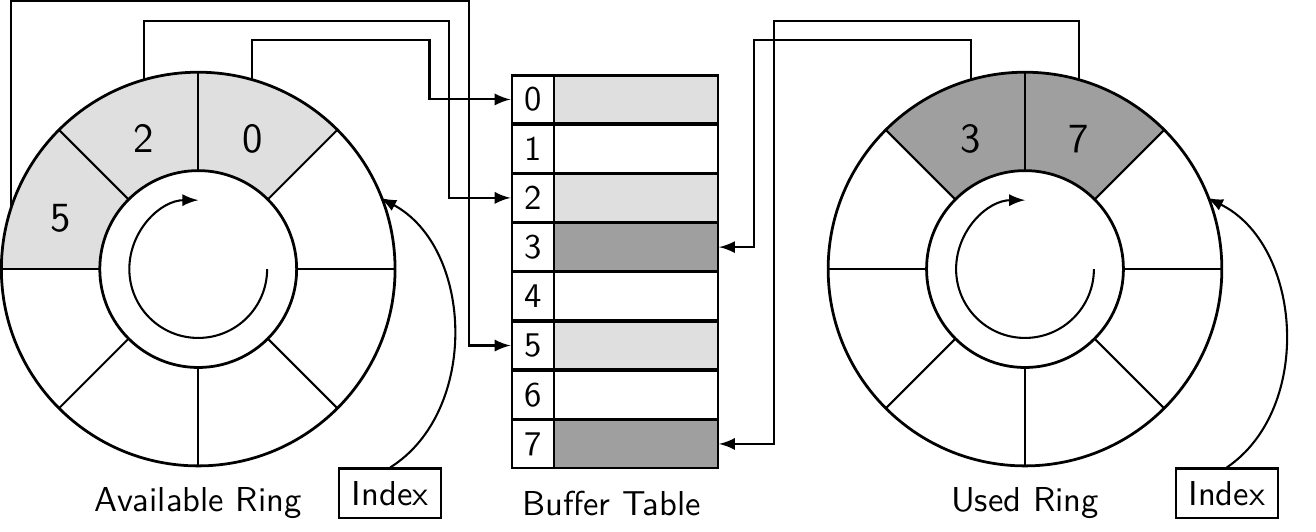}
\caption{Overview of a Virtqueue. Descriptor table contains physical addresses, the queues indices into the descriptor table.}
\vspace{-0.3cm}
\label{fig:virtqueue}
\vspace{-0.2cm}
\end{figure}

\href{https://github.com/emmericp/ixy/blob/b1cfa2240655f2644f7218abad3141236168f005/src/driver/virtio.c#L275}{\texttt{virtio\_legacy\_init()}} resets and configures a VirtIO device.
It negotiates the VirtIO version and features to use. 
See specification Section 5.1.3 and 5.1.5 for the available feature flags and initialization steps.

VirtIO supports three different types of queues called Virtqueues: receive, transmit, and command queues. 
The queue sizes are controlled by the device and are fixed to 256 entries for legacy devices.
Setup works the same as in the ixgbe driver: DMA memory for shared structures is allocated and passed to the device via a control register.
Contrary to queues in ixgbe, a Virtqueue internally consists of a descriptor table and two rings: the \emph{available} and \emph{used} rings.
While the table holds the complete descriptors with pointers to the physical addresses and length information of buffers, the rings only contain indices for this table as shown in Figure~\ref{fig:virtqueue}.
To supply a device with new buffers, the driver first adds new descriptors into free slots in the descriptor table and then enqueues the slot indices into the \emph{available} ring by advancing its head.
Conversely, a device picks up new descriptor indices from this ring, takes ownership of them and then signals completion by enqueuing the indices into the \emph{used} ring, where the driver finalizes the operation by clearing the descriptor from the table.
The queue indices are maintained in DMA memory instead of in registers like in the ixgbe implementation.
Therefore, the device needs to be informed about all modifications to queues, this is done by writing the queue ID into a control register in the IO port memory region.
Our driver also implements batching here to avoid unnecessary updates.
This process is the same for sending and receiving packets.
Our implementations are in \href{https://github.com/emmericp/ixy/blob/b1cfa2240655f2644f7218abad3141236168f005/src/driver/virtio.c#L49}{\texttt{virtio\_legacy\_setup\_tx}}/\href{https://github.com/emmericp/ixy/blob/b1cfa2240655f2644f7218abad3141236168f005/src/driver/virtio.c#L223}{\texttt{rx\_queue()}}.

The command queue is a transmit queue that is used to control most features of the device instead of via registers.
For example, enabling or disabling promiscuous mode in \href{https://github.com/emmericp/ixy/blob/b1cfa2240655f2644f7218abad3141236168f005/src/driver/virtio.c#L192}{\texttt{virtio\_legacy\_set\_promiscuous()}}
is done by sending a command packet with the appropriate flags through this queue.
See specification Section 5.1.6.5 for details on the command queue.
This way of controlling devices is not unique to virtual devices.
For example, the Intel XL710 40\,Gbit/s configures most features by sending messages to the firmware running on the device~\cite{xl710}.

\subsection{Packet Handling}
Packet transmission in
\href{https://github.com/emmericp/ixy/blob/b1cfa2240655f2644f7218abad3141236168f005/src/driver/virtio.c#L433}{\texttt{virtio\_tx\_batch()}}
and reception in
\href{https://github.com/emmericp/ixy/blob/b1cfa2240655f2644f7218abad3141236168f005/src/driver/virtio.c#L369}{\texttt{virtio\_rx\_batch()}}
works similar to the ixgbe driver.
The big difference to ixgbe is passing of metadata and offloading information.
Virtqueues are not only used for VirtIO network devices, but for other VirtIO devices as well.
DMA descriptors do not contain information specific for network devices.
Packets going through Virtqueues have this information prepended in an extra header in the DMA buffer.

This means that the transmit function needs to prepend an additional header to each packet, and our goal to support device-agnostic applications means that the application cannot know about this requirement when allocating memory.
ixy handles this by placing this extra header in front of the packet as VirtIO DMA requires no alignment on cache lines.
Our packet buffers already contain metadata before the actual packet to track the physical address and the owning memory pool.
Packet data starts at an offset of one cache line (64 byte) in the packet buffer, due to alignment requirements of other NICs.
This metadata cache line has enough space to accommodate the additional VirtIO header, we have explicitly marked this available area as \emph{head room} for drivers requiring this.
Our receive function offsets the address in the DMA descriptor by the appropriate amount to receive the extra header in the head room.
The user's ixy application treats the metadata header as opaque data.

\subsection{VirtIO Performance}
Performance with VirtIO is dominated by the implementation of the virtual device, i.e., the hypervisor, and not the driver in the virtual machine.
It is also possible to implement the hypervisor part of VirtIO, i.e., the device, in a separate user space application via the Vhost-user interface of qemu~\cite{vhost-user}.
Implementations of this exist in both Snabb und DPDK.
We only present baseline performance measurements running on kvm with Open vSwitch and in VirtualBox, because we are not interested in getting the fastest possible result, but reproducible results in a realistic environment.
Optimizations on the hypervisor are out of scope for this paper.

Running ixy in qemu-kvm 2.7.1 on a Xeon E3-1230 V2 CPU clocked at 3.30 GHz yields a performance of only 0.94\,Mpps for the \texttt{ixy-pktgen} application and 0.36\,Mpps for \texttt{ixy-fwd}.
DPDK is only marginally faster on the same setup: it manages to forward 0.4\,Mpps, these slow speeds are not unexpected on unoptimized hypervisors~\cite{EmmRauGal17}.
Performance is limited by packet rate, not data rate. Profiling with 1514\,byte packets yield near identical results with a forwarding rate of 4.8\,Gbit/s.
VMs often send even larger packets with an offloading feature known as generic segmentation offloading offered by VirtIO to achieve higher rates.
Profiling on the hypervisor shows that the interconnect is the bottleneck.
It fully utilizes one core to forward packets with Open vSwitch 2.6 through the kernel to the second VM.
Performance is even worse on VirtualBox 5.2 in our Vagrant setup~\cite{ixy-vagrant}.
It merely achieves 0.05\,Mpps on Linux with a 3.3\,GHz Xeon E3 CPU and 0.06\,Mpps on macOS with a 2.3\,GHz Core i7 CPU (606\,Mbit/s with 1514\,byte packets).
DPDK achieves 0.08\,Mpps on the macOS setup.
Profiling within the VM shows that over 99\% of the CPU time is spent on an x86 \texttt{OUT} IO instruction to communicate with the virtual device.

\section{Conclusions}
We discussed how to build a user space driver for NICs of the ixgbe family which are commonly found in servers and for virtual VirtIO NICs. 
Our performance evaluation offers some unprecedented looks into performance of user space drivers.
ixy allows us to assess effects of individual optimizations, like DMA buffers allocated on huge pages, in isolation.
Our driver allowed for a simple port to normal-sized pages without IOMMU, this would be significant change in other frameworks\footnote{DPDK offers \texttt{--no-huge}, but this setting is incompatible with most DMA drivers due to the aforementioned safety issues.}.
Not everyone has access to servers with 10\,Gbit/s NICs to play around with driver development.
However, everyone can build a VM setup to test ixy with our VirtIO driver.
Our Vagrant setup is the simplest way to run ixy in a VM, instructions are in our repository~\cite{ixy-vagrant}.

\subsection*{Drivers in High-Level Languages}
The implementation presented here is written in C with the hope to be readable by everyone.
But there is nothing tying user space drivers to traditional systems programming languages.
We also implemented the same driver in Rust, Go, C\#, Java, OCaml, Haskell, Swift, JavaScript, and Python to compare these languages for user space drivers~\cite{ixy-languages-paper}.

\subsection*{Reproducible Research}

Scripts used for the evaluation and our DPDK forwarding application used for comparison are available in~\cite{ixy-scripts}.
We used commit \emph{df1cddbb} of ixy for the evaluation of ixgbe and virtio, commit \emph{a0f618d} on a branch~\cite{ixy-smallpage} for the normal sized pages.
Most results were obtained on an Intel Xeon E5-2620 v3 2.4 GHz CPU running Debian 9 (kernel 4.9) with a dual port Intel 82599ES NIC.
The NUMA results where obtained on a system with two Intel Xeon E5-2630 v4 2.2 GHz CPUs with Intel X550 NICs.
Turboboost, Hyper-Threading, and power-saving features were disabled.
VirtIO results were obtained on various systems and hypervisors as described in the evaluation section.
All loads where generated with MoonGen~\cite{moongen} and its \texttt{l2-load-latency.lua} script. 

\subsection*{Acknowledgments}
This work was supported by the German-French Academy for the Industry of the Future.
We would like to thank Simon Ellmann, Masanori Misono, and Boris-Chengbiao Zhou for valuable contributions to ixy and/or this paper.

\appto\UrlNoBreaks{\do\:}
\vspace{1ex}
\pagebreak
\bibliographystyle{ACM-Reference-Format}
\renewcommand*{\bibfont}{\fontsize{10}{12}\selectfont}
{\large\bibliography{lit}}


\begin{thebibliography}{46}


\ifx \showCODEN    \undefined \def \showCODEN     #1{\unskip}     \fi
\ifx \showDOI      \undefined \def \showDOI       #1{#1}\fi
\ifx \showISBNx    \undefined \def \showISBNx     #1{\unskip}     \fi
\ifx \showISBNxiii \undefined \def \showISBNxiii  #1{\unskip}     \fi
\ifx \showISSN     \undefined \def \showISSN      #1{\unskip}     \fi
\ifx \showLCCN     \undefined \def \showLCCN      #1{\unskip}     \fi
\ifx \shownote     \undefined \def \shownote      #1{#1}          \fi
\ifx \showarticletitle \undefined \def \showarticletitle #1{#1}   \fi
\ifx \showURL      \undefined \def \showURL       {\relax}        \fi
\providecommand\bibfield[2]{#2}
\providecommand\bibinfo[2]{#2}
\providecommand\natexlab[1]{#1}
\providecommand\showeprint[2][]{arXiv:#2}

\bibitem[\protect\citeauthoryear{Bainbridge and Maxwell}{Bainbridge and
  Maxwell}{2015}]%
        {redhattuning}
\bibfield{author}{\bibinfo{person}{Jamie Bainbridge} {and} \bibinfo{person}{Jon
  Maxwell}.} \bibinfo{year}{2015}\natexlab{}.
\newblock \showarticletitle{{Red Hat Enterprise Linux Network Performance
  Tuning Guide}}.
\newblock \bibinfo{journal}{{\em {Red Hat Documentation}\/}}
  (\bibinfo{date}{March} \bibinfo{year}{2015}).
\newblock
\newblock
\shownote{\texttt{\url{https://access.redhat.com/sites/default/files/attachments/20150325_network_performance_tuning.pdf}}.}


\bibitem[\protect\citeauthoryear{Barbette, Soldani, and Mathy}{Barbette
  et~al\mbox{.}}{2015}]%
        {fastclick}
\bibfield{author}{\bibinfo{person}{Tom Barbette}, \bibinfo{person}{Cyril
  Soldani}, {and} \bibinfo{person}{Laurent Mathy}.}
  \bibinfo{year}{2015}\natexlab{}.
\newblock \showarticletitle{Fast userspace packet processing}. In
  \bibinfo{booktitle}{{\em ACM/IEEE ANCS}}.
\newblock


\bibitem[\protect\citeauthoryear{Bertin}{Bertin}{2015}]%
        {cloudflare-netmap}
\bibfield{author}{\bibinfo{person}{Gilberto Bertin}.}
  \bibinfo{year}{2015}\natexlab{}.
\newblock \bibinfo{title}{{Single RX queue kernel bypass in Netmap for high
  packet rate networking}}.
\newblock   (\bibinfo{date}{Oct.} \bibinfo{year}{2015}).
\newblock
\newblock
\shownote{\texttt{\url{https://blog.cloudflare.com/single-rx-queue-kernel-bypass-with-netmap/}}.}


\bibitem[\protect\citeauthoryear{Bonelli, Giordano, and Procissi}{Bonelli
  et~al\mbox{.}}{2016}]%
        {pfq}
\bibfield{author}{\bibinfo{person}{N. Bonelli}, \bibinfo{person}{S. Giordano},
  {and} \bibinfo{person}{G. Procissi}.} \bibinfo{year}{2016}\natexlab{}.
\newblock \showarticletitle{Network Traffic Processing With PFQ}.
\newblock \bibinfo{journal}{{\em IEEE Journal on Selected Areas in
  Communications\/}} \bibinfo{volume}{34}, \bibinfo{number}{6}
  (\bibinfo{date}{June} \bibinfo{year}{2016}), \bibinfo{pages}{1819--1833}.
\newblock
\showISSN{0733-8716}
\showDOI{%
\url{https://doi.org/10.1109/JSAC.2016.2558998}}


\bibitem[\protect\citeauthoryear{{DPDK Project}}{{DPDK Project}}{2019a}]%
        {dpdk-nics}
\bibfield{author}{\bibinfo{person}{{DPDK Project}}.}
  \bibinfo{year}{2019}\natexlab{a}.
\newblock \bibinfo{title}{{DPDK: Supported NICs}}.
\newblock   (\bibinfo{year}{2019}).
\newblock
\newblock
\shownote{\texttt{\url{http://dpdk.org/doc/nics}}.}


\bibitem[\protect\citeauthoryear{{DPDK Project}}{{DPDK Project}}{2019b}]%
        {dpdk-support-matrix}
\bibfield{author}{\bibinfo{person}{{DPDK Project}}.}
  \bibinfo{year}{2019}\natexlab{b}.
\newblock \bibinfo{title}{{DPDK User Guide: Overview of Networking Drivers}}.
\newblock   (\bibinfo{year}{2019}).
\newblock
\newblock
\shownote{\texttt{\url{http://dpdk.org/doc/guides/nics/overview.html}}.}


\bibitem[\protect\citeauthoryear{{DPDK Project}}{{DPDK Project}}{2019c}]%
        {dpdk}
\bibfield{author}{\bibinfo{person}{{DPDK Project}}.}
  \bibinfo{year}{2019}\natexlab{c}.
\newblock \bibinfo{title}{{DPDK Website}}.
\newblock   (\bibinfo{year}{2019}).
\newblock
\newblock
\shownote{\texttt{\url{http://dpdk.org/}}.}


\bibitem[\protect\citeauthoryear{Emmerich, Ellmann, Bonk, Egger,
  Sánchez-Torija, Günzel, Luzio, Obada, Stadlmeier, Voit, and Carle}{Emmerich
  et~al\mbox{.}}{2019}]%
        {ixy-languages-paper}
\bibfield{author}{\bibinfo{person}{Paul Emmerich}, \bibinfo{person}{Simon
  Ellmann}, \bibinfo{person}{Fabian Bonk}, \bibinfo{person}{Alex Egger},
  \bibinfo{person}{Esaú~García Sánchez-Torija}, \bibinfo{person}{Thomas
  Günzel}, \bibinfo{person}{Sebastian~Di Luzio}, \bibinfo{person}{Alexandru
  Obada}, \bibinfo{person}{Maximilian Stadlmeier}, \bibinfo{person}{Sebastian
  Voit}, {and} \bibinfo{person}{Georg Carle}.} \bibinfo{year}{2019}\natexlab{}.
\newblock \showarticletitle{{The Case for Writing Network Drivers in High-Level
  Programming Languages}}. In \bibinfo{booktitle}{{\em ACM/IEEE Symposium on
  Architectures for Networking and Communications Systems (ANCS 2019)}}.
\newblock


\bibitem[\protect\citeauthoryear{{Emmerich}, {Gallenm{\"u}ller}, {Raumer},
  {Wohlfart}, and {Carle}}{{Emmerich} et~al\mbox{.}}{2015}]%
        {moongen}
\bibfield{author}{\bibinfo{person}{Paul {Emmerich}}, \bibinfo{person}{Sebastian
  {Gallenm{\"u}ller}}, \bibinfo{person}{Daniel {Raumer}},
  \bibinfo{person}{Florian {Wohlfart}}, {and} \bibinfo{person}{Georg {Carle}}.}
  \bibinfo{year}{2015}\natexlab{}.
\newblock \showarticletitle{{MoonGen: A Scriptable High-Speed Packet
  Generator}}. In \bibinfo{booktitle}{{\em Internet Measurement Conference 2015
  (IMC'15)}}. \bibinfo{address}{Tokyo, Japan}.
\newblock


\bibitem[\protect\citeauthoryear{Emmerich, Raumer, Gallenm\"{u}ller, Wohlfart,
  and Carle}{Emmerich et~al\mbox{.}}{2017}]%
        {EmmRauGal17}
\bibfield{author}{\bibinfo{person}{Paul Emmerich}, \bibinfo{person}{Daniel
  Raumer}, \bibinfo{person}{Sebastian Gallenm\"{u}ller},
  \bibinfo{person}{Florian Wohlfart}, {and} \bibinfo{person}{Georg Carle}.}
  \bibinfo{year}{2017}\natexlab{}.
\newblock \showarticletitle{{Throughput and Latency of Virtual Switching with
  Open vSwitch: A Quantitative Analysis}}.
\newblock \bibinfo{journal}{{\em Journal of Network and Systems Management\/}}
  (\bibinfo{date}{July} \bibinfo{year}{2017}).
\newblock
\showDOI{%
\url{https://doi.org/10.1007/s10922-017-9417-0}}


\bibitem[\protect\citeauthoryear{{FreeBSD Project}}{{FreeBSD Project}}{2017}]%
        {netmap-offload}
\bibfield{author}{\bibinfo{person}{{FreeBSD Project}}.}
  \bibinfo{year}{2017}\natexlab{}.
\newblock \showarticletitle{{NETMAP(4)}}. In \bibinfo{booktitle}{{\em {FreeBSD
  Kernel Interfaces Manual}}}. {FreeBSD 11.1-RELEASE}.
\newblock


\bibitem[\protect\citeauthoryear{Gallenmüller, Emmerich, Wohlfart, Raumer, and
  Carle}{Gallenmüller et~al\mbox{.}}{2015}]%
        {gallenmueller2015comparison}
\bibfield{author}{\bibinfo{person}{Sebastian Gallenmüller},
  \bibinfo{person}{Paul Emmerich}, \bibinfo{person}{Florian Wohlfart},
  \bibinfo{person}{Daniel Raumer}, {and} \bibinfo{person}{Georg Carle}.}
  \bibinfo{year}{2015}\natexlab{}.
\newblock \showarticletitle{{Comparison of Frameworks for High-Performance
  Packet IO}}. In \bibinfo{booktitle}{{\em Architectures for Networking and
  Communications Systems (ANCS)}}. ACM, \bibinfo{address}{Oakland, CA},
  \bibinfo{pages}{29--38}.
\newblock


\bibitem[\protect\citeauthoryear{Gettys and Nichols}{Gettys and
  Nichols}{2011}]%
        {gettys2011bufferbloat}
\bibfield{author}{\bibinfo{person}{Jim Gettys} {and} \bibinfo{person}{Kathleen
  Nichols}.} \bibinfo{year}{2011}\natexlab{}.
\newblock \showarticletitle{{Bufferbloat: Dark buffers in the internet}}.
\newblock \bibinfo{journal}{{\em Queue\/}} \bibinfo{volume}{9},
  \bibinfo{number}{11} (\bibinfo{year}{2011}), \bibinfo{pages}{40}.
\newblock


\bibitem[\protect\citeauthoryear{{Gilberto Bertin}}{{Gilberto Bertin}}{2017}]%
        {xdp-ddos}
\bibfield{author}{\bibinfo{person}{{Gilberto Bertin}}.}
  \bibinfo{year}{2017}\natexlab{}.
\newblock \showarticletitle{{XDP in practice: integrating XDP into our DDoS
  mitigation pipeline}}. In \bibinfo{booktitle}{{\em Netdev 2.1, The Technical
  Conference on Linux Networking}}.
\newblock


\bibitem[\protect\citeauthoryear{{Gorrie, L et al.}}{{Gorrie, L et
  al.}}{2019}]%
        {snabb}
\bibfield{author}{\bibinfo{person}{{Gorrie, L et al.}}}
  \bibinfo{year}{2019}\natexlab{}.
\newblock \bibinfo{title}{{ Snabb: Simple and fast packet networking}}.
\newblock   (\bibinfo{year}{2019}).
\newblock
\newblock
\shownote{\texttt{\url{https://github.com/snabbco/snabb}}.}


\bibitem[\protect\citeauthoryear{Haardt}{Haardt}{1993}]%
        {ioperm}
\bibfield{author}{\bibinfo{person}{Michael Haardt}.}
  \bibinfo{year}{1993}\natexlab{}.
\newblock \showarticletitle{{ioperm(2)}}. In \bibinfo{booktitle}{{\em Linux
  Programmer's Manual}}.
\newblock


\bibitem[\protect\citeauthoryear{{HashiCorp}}{{HashiCorp}}{2019}]%
        {vagrant}
\bibfield{author}{\bibinfo{person}{{HashiCorp}}.}
  \bibinfo{year}{2019}\natexlab{}.
\newblock \bibinfo{title}{{Vagrant website}}.
\newblock   (\bibinfo{year}{2019}).
\newblock
\newblock
\shownote{\texttt{\url{https://www.vagrantup.com/}}.}


\bibitem[\protect\citeauthoryear{Intel}{Intel}{2012}]%
        {ddio}
\bibfield{author}{\bibinfo{person}{Intel}.} \bibinfo{year}{2012}\natexlab{}.
\newblock \showarticletitle{{Intel Data Direct I/O Technology (Intel DDIO): A
  Primer}}.
\newblock  (\bibinfo{year}{2012}).
\newblock
\newblock
\shownote{\texttt{\url{https://www.intel.com/content/www/us/en/io/data-direct-i-o-technology-brief.html}}.}


\bibitem[\protect\citeauthoryear{Intel}{Intel}{2014}]%
        {xl710}
\bibfield{author}{\bibinfo{person}{Intel}.} \bibinfo{year}{2014}\natexlab{}.
\newblock \showarticletitle{{Intel Ethernet Controller XL710 Datasheet Rev.
  2.1}}.
\newblock


\bibitem[\protect\citeauthoryear{Intel}{Intel}{2016}]%
        {82599}
\bibfield{author}{\bibinfo{person}{Intel}.} \bibinfo{year}{2016}\natexlab{}.
\newblock \showarticletitle{{Intel 82599 10 GbE Controller Datasheet Rev.
  3.3}}.
\newblock


\bibitem[\protect\citeauthoryear{Intel}{Intel}{2019}]%
        {hugepageperformance1}
\bibfield{author}{\bibinfo{person}{Intel}.} \bibinfo{year}{2019}\natexlab{}.
\newblock \bibinfo{title}{{DPDK Getting Started Guide for Linux}}.
\newblock   (\bibinfo{year}{2019}).
\newblock
\newblock
\shownote{\texttt{\url{http://dpdk.org/doc/guides/linux_gsg/sys_reqs.html}}.}


\bibitem[\protect\citeauthoryear{{IO Visor Project}}{{IO Visor
  Project}}{2019a}]%
        {xdp-drivers}
\bibfield{author}{\bibinfo{person}{{IO Visor Project}}.}
  \bibinfo{year}{2019}\natexlab{a}.
\newblock \bibinfo{title}{{BPF and XDP Features by Kernel Version}}.
\newblock   (\bibinfo{year}{2019}).
\newblock
\newblock
\shownote{\texttt{\url{https://github.com/iovisor/bcc/blob/master/docs/kernel-versions.md\#xdp}}.}


\bibitem[\protect\citeauthoryear{{IO Visor Project}}{{IO Visor
  Project}}{2019b}]%
        {xdp}
\bibfield{author}{\bibinfo{person}{{IO Visor Project}}.}
  \bibinfo{year}{2019}\natexlab{b}.
\newblock \bibinfo{title}{{Introduction to XDP}}.
\newblock   (\bibinfo{year}{2019}).
\newblock
\newblock
\shownote{\texttt{\url{https://www.iovisor.org/technology/xdp}}.}


\bibitem[\protect\citeauthoryear{{Jim Thompson}}{{Jim Thompson}}{2017}]%
        {pfsensedpdk}
\bibfield{author}{\bibinfo{person}{{Jim Thompson}}.}
  \bibinfo{year}{2017}\natexlab{}.
\newblock \showarticletitle{{DPDK, VPP \& pfSense 3.0}}. In
  \bibinfo{booktitle}{{\em {DPDK Summit Userspace}}}.
\newblock


\bibitem[\protect\citeauthoryear{{Jonathan Corbet}}{{Jonathan Corbet}}{2017}]%
        {snabb-design-goals}
\bibfield{author}{\bibinfo{person}{{Jonathan Corbet}}.}
  \bibinfo{year}{2017}\natexlab{}.
\newblock \showarticletitle{{User-space networking with Snabb}}. In
  \bibinfo{booktitle}{{\em LWN.net}}.
\newblock


\bibitem[\protect\citeauthoryear{Kerrisk}{Kerrisk}{2004}]%
        {mlock}
\bibfield{author}{\bibinfo{person}{Michael Kerrisk}.}
  \bibinfo{year}{2004}\natexlab{}.
\newblock \showarticletitle{{mlock(2)}}. In \bibinfo{booktitle}{{\em Linux
  Programmer's Manual}}.
\newblock


\bibitem[\protect\citeauthoryear{{Linux Foundation}}{{Linux
  Foundation}}{2017}]%
        {dpdk-linux-foundation}
\bibfield{author}{\bibinfo{person}{{Linux Foundation}}.}
  \bibinfo{year}{2017}\natexlab{}.
\newblock \bibinfo{title}{{Networking Industry Leaders Join Forces to Expand
  New Open Source Community to Drive Development of the DPDK Project}}.
\newblock   (\bibinfo{date}{April} \bibinfo{year}{2017}).
\newblock
\newblock
\shownote{Press release.}


\bibitem[\protect\citeauthoryear{{Linux Kernel Documentation}}{{Linux Kernel
  Documentation}}{2019a}]%
        {pagemigration}
\bibfield{author}{\bibinfo{person}{{Linux Kernel Documentation}}.}
  \bibinfo{year}{2019}\natexlab{a}.
\newblock \bibinfo{title}{{Page migration}}.
\newblock   (\bibinfo{year}{2019}).
\newblock
\newblock
\shownote{\texttt{\url{https://www.kernel.org/doc/Documentation/vm/page_migration}}.}


\bibitem[\protect\citeauthoryear{{Linux Kernel Documentation}}{{Linux Kernel
  Documentation}}{2019b}]%
        {vfio}
\bibfield{author}{\bibinfo{person}{{Linux Kernel Documentation}}.}
  \bibinfo{year}{2019}\natexlab{b}.
\newblock \bibinfo{title}{{VFIO - Virtual Function I/O}}.
\newblock   (\bibinfo{year}{2019}).
\newblock
\newblock
\shownote{\texttt{\url{https://www.kernel.org/doc/Documentation/vfio.txt}}.}


\bibitem[\protect\citeauthoryear{{Maximilian Pudelko}}{{Maximilian
  Pudelko}}{2019a}]%
        {ixy-smallpage}
\bibfield{author}{\bibinfo{person}{{Maximilian Pudelko}}.}
  \bibinfo{year}{2019}\natexlab{a}.
\newblock \bibinfo{title}{{ixy - DMA allocator on normal-sized pages}}.
\newblock   (\bibinfo{year}{2019}).
\newblock
\newblock
\shownote{\texttt{\url{https://github.com/pudelkoM/ixy/tree/contiguous-pages}}.}


\bibitem[\protect\citeauthoryear{{Maximilian Pudelko}}{{Maximilian
  Pudelko}}{2019b}]%
        {ixy-headptr-writeback}
\bibfield{author}{\bibinfo{person}{{Maximilian Pudelko}}.}
  \bibinfo{year}{2019}\natexlab{b}.
\newblock \bibinfo{title}{{ixy - head pointer writeback implementation}}.
\newblock   (\bibinfo{year}{2019}).
\newblock
\newblock
\shownote{\texttt{\url{https://github.com/pudelkoM/ixy/tree/head-pointer-writeback}}.}


\bibitem[\protect\citeauthoryear{Morris, Kohler, Jannotti, and
  Frans~Kaashoek}{Morris et~al\mbox{.}}{1999}]%
        {click}
\bibfield{author}{\bibinfo{person}{Robert Morris}, \bibinfo{person}{Eddie
  Kohler}, \bibinfo{person}{John Jannotti}, {and} \bibinfo{person}{M
  Frans~Kaashoek}.} \bibinfo{year}{1999}\natexlab{}.
\newblock \showarticletitle{The Click modular router}. In
  \bibinfo{booktitle}{{\em Operating Systems Review - SIGOPS}},
  Vol.~\bibinfo{volume}{33}. \bibinfo{pages}{217--231}.
\newblock


\bibitem[\protect\citeauthoryear{Neugebauer, Antichi, Zazo, Audzevich,
  L{\'o}pez-Buedo, and Moore}{Neugebauer et~al\mbox{.}}{2018}]%
        {pciebench}
\bibfield{author}{\bibinfo{person}{Rolf Neugebauer}, \bibinfo{person}{Gianni
  Antichi}, \bibinfo{person}{Jos{\'e}~Fernando Zazo}, \bibinfo{person}{Yury
  Audzevich}, \bibinfo{person}{Sergio L{\'o}pez-Buedo}, {and}
  \bibinfo{person}{Andrew~W Moore}.} \bibinfo{year}{2018}\natexlab{}.
\newblock \showarticletitle{Understanding PCIe performance for end host
  networking}. In \bibinfo{booktitle}{{\em SIGCOMM 2018}}. ACM,
  \bibinfo{pages}{327--341}.
\newblock


\bibitem[\protect\citeauthoryear{ntop}{ntop}{2014}]%
        {pfring}
\bibfield{author}{\bibinfo{person}{ntop}.} \bibinfo{year}{2014}\natexlab{}.
\newblock \bibinfo{title}{{PF\_RING ZC (Zero Copy)}}.
\newblock   (\bibinfo{year}{2014}).
\newblock
\newblock
\shownote{\texttt{\url{http://www.ntop.org/products/packet-capture/pf_ring/pf_ring-zc-zero-copy/}}.}


\bibitem[\protect\citeauthoryear{{OASIS VIRTIO TC}}{{OASIS VIRTIO TC}}{2016}]%
        {virtiospec}
\bibfield{author}{\bibinfo{person}{{OASIS VIRTIO TC}}.}
  \bibinfo{year}{2016}\natexlab{}.
\newblock \bibinfo{title}{{Virtual I/O Device (VIRTIO) Version 1.0}}.
\newblock   (\bibinfo{date}{March} \bibinfo{year}{2016}).
\newblock
\newblock
\shownote{\texttt{\url{http://docs.oasis-open.org/virtio/virtio/v1.0/virtio-v1.0.pdf}}.}


\bibitem[\protect\citeauthoryear{{Open vSwitch Project}}{{Open vSwitch
  Project}}{2019}]%
        {ovs-dpdk}
\bibfield{author}{\bibinfo{person}{{Open vSwitch Project}}.}
  \bibinfo{year}{2019}\natexlab{}.
\newblock \bibinfo{title}{{Open vSwitch with DPDK}}.
\newblock   (\bibinfo{year}{2019}).
\newblock
\newblock
\shownote{\texttt{\url{http://docs.openvswitch.org/en/latest/intro/install/dpdk/}}.}


\bibitem[\protect\citeauthoryear{{Paul Emmerich}}{{Paul Emmerich}}{2019}]%
        {ixy-vagrant}
\bibfield{author}{\bibinfo{person}{{Paul Emmerich}}.}
  \bibinfo{year}{2019}\natexlab{}.
\newblock \bibinfo{title}{{ixy Vagrant setup}}.
\newblock   (\bibinfo{year}{2019}).
\newblock
\newblock
\shownote{\texttt{\url{https://github.com/emmericp/ixy/tree/master/vagrant}}.}


\bibitem[\protect\citeauthoryear{{Paul Emmerich et al.}}{{Paul Emmerich et
  al.}}{2019}]%
        {ixy}
\bibfield{author}{\bibinfo{person}{{Paul Emmerich et al.}}}
  \bibinfo{year}{2019}\natexlab{}.
\newblock \bibinfo{title}{{ixy code}}.
\newblock   (\bibinfo{year}{2019}).
\newblock
\newblock
\shownote{\texttt{\url{https://github.com/emmericp/ixy}}.}


\bibitem[\protect\citeauthoryear{{Paul Emmerich, Simon Bauer}}{{Paul Emmerich,
  Simon Bauer}}{2019}]%
        {ixy-scripts}
\bibfield{author}{\bibinfo{person}{{Paul Emmerich, Simon Bauer}}.}
  \bibinfo{year}{2019}\natexlab{}.
\newblock \bibinfo{title}{{Scripts used for the performance evaluation}}.
\newblock   (\bibinfo{year}{2019}).
\newblock
\newblock
\shownote{\texttt{\url{https://github.com/emmericp/ixy-perf-measurements}}.}


\bibitem[\protect\citeauthoryear{Pfaff, Pettit, Koponen, Jackson, Zhou,
  Rajahalme, Gross, Wang, Stringer, Shelar, Amidon, and Casado}{Pfaff
  et~al\mbox{.}}{2015}]%
        {ovs}
\bibfield{author}{\bibinfo{person}{Ben Pfaff}, \bibinfo{person}{Justin Pettit},
  \bibinfo{person}{Teemu Koponen}, \bibinfo{person}{Ethan Jackson},
  \bibinfo{person}{Andy Zhou}, \bibinfo{person}{Jarno Rajahalme},
  \bibinfo{person}{Jesse Gross}, \bibinfo{person}{Alex Wang},
  \bibinfo{person}{Joe Stringer}, \bibinfo{person}{Pravin Shelar},
  \bibinfo{person}{Keith Amidon}, {and} \bibinfo{person}{Martin Casado}.}
  \bibinfo{year}{2015}\natexlab{}.
\newblock \showarticletitle{The Design and Implementation of Open vSwitch}. In
  \bibinfo{booktitle}{{\em 12th {USENIX} Symposium on Networked Systems Design
  and Implementation ({NSDI} 15)}}. \bibinfo{publisher}{{USENIX} Association},
  \bibinfo{address}{Oakland, CA}, \bibinfo{pages}{117--130}.
\newblock
\showISBNx{978-1-931971-218}


\bibitem[\protect\citeauthoryear{Rizzo}{Rizzo}{2012}]%
        {netmap}
\bibfield{author}{\bibinfo{person}{Luigi Rizzo}.}
  \bibinfo{year}{2012}\natexlab{}.
\newblock \showarticletitle{{netmap: A Novel Framework for Fast Packet I/O.}}.
  In \bibinfo{booktitle}{{\em {USENIX Annual Technical Conference}}}.
  \bibinfo{pages}{101--112}.
\newblock


\bibitem[\protect\citeauthoryear{{Snabb Project}}{{Snabb Project}}{2018}]%
        {hugepageperformance2}
\bibfield{author}{\bibinfo{person}{{Snabb Project}}.}
  \bibinfo{year}{2018}\natexlab{}.
\newblock \bibinfo{title}{{Tuning the performance of the lwaftr}}.
\newblock   (\bibinfo{year}{2018}).
\newblock
\newblock
\shownote{\texttt{\url{https://github.com/snabbco/snabb/blob/master/src/program/lwaftr/doc/performance.md}}.}


\bibitem[\protect\citeauthoryear{{Snort Project}}{{Snort Project}}{2015}]%
        {snortreadme}
\bibfield{author}{\bibinfo{person}{{Snort Project}}.}
  \bibinfo{year}{2015}\natexlab{}.
\newblock \bibinfo{title}{{Snort 3 User Manual}}.
\newblock   (\bibinfo{year}{2015}).
\newblock
\newblock
\shownote{\texttt{\url{https://www.snort.org/downloads/snortplus/snort_manual.pdf}}.}


\bibitem[\protect\citeauthoryear{{Solarflare}}{{Solarflare}}{2019}]%
        {openonload}
\bibfield{author}{\bibinfo{person}{{Solarflare}}.}
  \bibinfo{year}{2019}\natexlab{}.
\newblock \bibinfo{title}{{OpenOnload Website}}.
\newblock   (\bibinfo{year}{2019}).
\newblock
\newblock
\shownote{\texttt{\url{http://www.openonload.org/}}.}


\bibitem[\protect\citeauthoryear{{Virtual Open Systems Sarl.}}{{Virtual Open
  Systems Sarl.}}{2014}]%
        {vhost-user}
\bibfield{author}{\bibinfo{person}{{Virtual Open Systems Sarl.}}}
  \bibinfo{year}{2014}\natexlab{}.
\newblock \bibinfo{title}{{Vhost-user Protocol}}.
\newblock   (\bibinfo{year}{2014}).
\newblock
\newblock
\shownote{\texttt{\url{https://github.com/qemu/qemu/blob/stable-2.10/docs/interop/vhost-user.txt}}.}


\bibitem[\protect\citeauthoryear{Yasukata, Honda, Santry, and Eggert}{Yasukata
  et~al\mbox{.}}{2016}]%
        {stackmap}
\bibfield{author}{\bibinfo{person}{Kenichi Yasukata}, \bibinfo{person}{Michio
  Honda}, \bibinfo{person}{Douglas Santry}, {and} \bibinfo{person}{Lars
  Eggert}.} \bibinfo{year}{2016}\natexlab{}.
\newblock \showarticletitle{{StackMap: Low-Latency Networking with the {OS}
  Stack and Dedicated NICs}}. In \bibinfo{booktitle}{{\em 2016 {USENIX} Annual
  Technical Conference ({USENIX} {ATC} 16)}}. \bibinfo{publisher}{{USENIX}
  Association}, \bibinfo{address}{Denver, CO}, \bibinfo{pages}{43--56}.
\newblock
\showISBNx{978-1-931971-30-0}


\end{thebibliography}

\end{document}